\documentclass[12pt]{article}
\pdfoutput=1

\usepackage{amsmath,amssymb,amsfonts,epsfig,cite,setspace,bigstrut,framed}
\usepackage[all]{xy}
\usepackage{color}
\usepackage{pifont}

\makeatletter \@addtoreset{equation}{section} \makeatother
\renewcommand{\theequation}{\thesection.\arabic{equation}}

\begin{document}

\vskip 0.25in

\newcommand{\todo}[1]{{\bf\color{blue} !! #1 !!}\marginpar{\color{blue}$\Longleftarrow$}}
\newcommand{\nn}{\nonumber}
\newcommand{\comment}[1]{}
\newcommand\T{\rule{0pt}{2.6ex}}
\newcommand\B{\rule[-1.2ex]{0pt}{0pt}}

\newcommand{\CO}{{\cal O}}
\newcommand{\cI}{{\cal I}}
\newcommand{\cM}{{\cal M}}
\newcommand{\cW}{{\cal W}}
\newcommand{\cN}{{\cal N}}
\newcommand{\cR}{{\cal R}}
\newcommand{\cH}{{\cal H}}
\newcommand{\cK}{{\cal K}}
\newcommand{\cT}{{\cal T}}
\newcommand{\cZ}{{\cal Z}}
\newcommand{\cO}{{\cal O}}
\newcommand{\cQ}{{\cal Q}}
\newcommand{\cB}{{\cal B}}
\newcommand{\cC}{{\cal C}}
\newcommand{\cD}{{\cal D}}
\newcommand{\cE}{{\cal E}}
\newcommand{\cF}{{\cal F}}
\newcommand{\cG}{{\cal G}}
\newcommand{\cA}{{\cal A}}
\newcommand{\cX}{{\cal X}}
\newcommand{\IA}{\mathbb{A}}
\newcommand{\IP}{\mathbb{P}}
\newcommand{\IQ}{\mathbb{Q}}
\newcommand{\IH}{\mathbb{H}}
\newcommand{\IR}{\mathbb{R}}
\newcommand{\IC}{\mathbb{C}}
\newcommand{\IF}{\mathbb{F}}
\newcommand{\IV}{\mathbb{V}}
\newcommand{\II}{\mathbb{I}}
\newcommand{\IZ}{\mathbb{Z}}
\newcommand{\re}{{\rm Re}}
\newcommand{\im}{{\rm Im}}
\newcommand{\tr}{\mathop{\rm Tr}}
\newcommand{\ch}{{\rm ch}}
\newcommand{\rk}{{\rm rk}}
\newcommand{\ext}{{\rm Ext}}
\newcommand{\bi}{\begin{itemize}}
\newcommand{\ei}{\end{itemize}}
\newcommand{\beq}{\begin{equation}}
\newcommand{\eeq}{\end{equation}}
\newcommand{\bea}{\begin{eqnarray}}
\newcommand{\eea}{\end{eqnarray}}
\newcommand{\ba}{\begin{array}}
\newcommand{\ea}{\end{array}}

\newcommand{\CN}{{\cal N}}
\newcommand{\y}{{\mathbf y}}
\newcommand{\z}{{\mathbf z}}
\newcommand{\C}{\mathbb C}\newcommand{\R}{\mathbb R}
\newcommand{\CA}{\mathbb A}
\newcommand{\CP}{\mathbb P}
\newcommand{\cP}{\mathcal P}
\newcommand{\tmat}[1]{{\tiny \left(\begin{matrix} #1 \end{matrix}\right)}}
\newcommand{\mat}[1]{\left(\begin{matrix} #1 \end{matrix}\right)}
\newcommand{\diff}[2]{\frac{\partial #1}{\partial #2}}
\newcommand{\gen}[1]{\langle #1 \rangle}

\newtheorem{theorem}{\bf THEOREM}
\newtheorem{proposition}{\bf PROPOSITION}
\newtheorem{observation}{\bf OBSERVATION}

\def\theequation{\thesection.\arabic{equation}}
\newcommand{\setall}{
	\setcounter{equation}{0}
}
\renewcommand{\thefootnote}{\fnsymbol{footnote}}

\begin{titlepage}
\vfill
\begin{flushright}
{\tt\normalsize KIAS-P18005}\\
{\tt\normalsize arXiv:1801.05460}\\
\end{flushright}
\vfill

\begin{center}
{\Large\bf Holonomy Saddles  and Supersymmetry}

\vskip 1.5cm

Chiung Hwang\footnote{\tt chwang@kias.re.kr},
Sungjay Lee\footnote{\tt sjlee@kias.re.kr}, and
Piljin Yi\footnote{\tt piljin@kias.re.kr}
\vskip 5mm

{\it School of Physics,
Korea Institute for Advanced Study, Seoul 02455, Korea}

\end{center}
\vfill

\begin{abstract}

In gauge theories on a spacetime equipped with a circle,
the holonomy variables, living in the Cartan torus, play
special roles. With their periodic nature properly taken
into account, we find that a supersymmetric gauge theory
in $d$ dimensions  tends to reduce in the small radius
limit to a disjoint sum of multiple $(d-1)$ dimensional
theories at distinct holonomies, called $H$-saddles.
The phenomenon occurs regardless of the spacetime
dimensions, and here we explore such $H$-saddles
for $d=4$ $\cN=1$ theories on $T^2$ fibred over
$\Sigma_g$, in the limits of elongated $T^2$.
This naturally generates novel relationships between
4d and 3d partition functions, including ones between
4d and 3d Witten indices, and also leads us
to re-examine  recent studies of the Cardy exponents
and the Casimir energies and of their purported
connections to the 4d anomalies.
\end{abstract}

\vfill
\end{titlepage}

\tableofcontents
\renewcommand{\thefootnote}{\#\arabic{footnote}}
\setcounter{footnote}{0}

\newpage

\section{Gluing Gauge Theories across Dimensions}

Gauge theories in a spacetime with a circle admit
holonomy variables
as special degrees of freedom. With the spacetime
sufficiently noncompact, the infrared properties of
the theory is often characterized by the vacuum
expectation values (vev) of the Wilson line operator
\cite{Wilson:1974sk}, or the traced holonomy along the circle.

In many theories, the holonomy variables are not
exactly flat at the quantum level and the Wilson line
often serves as an order parameter.
For example, 4d $\cN=1$ pure $SU(N)$ Yang-Mills
on a large circle, or on a circle with supersymmetric
boudnary condition, are known to admit
$N$ distinct vacua, whose confining nature is
dictated by equally spaced eigenvalues of the holonomy,
hence a vanishing Wilson line expectation value.
If we replace the circle by a sufficiently
small thermal circle, with the anti-periodic boundary condition
on gauginos, the eigenvalues become clustered at the origin,
signalling a de-confined phase at high temperature as
evidenced by a non-vanishing Wilson line vev.

If supersymmetry is extended enough to ensure that these
variables correspond to genuine flat directions at
quantum level, compactification on the circle generates
an infinite number of superselection
sectors, labeled by the holonomy. A more typical situation
with minimal supersymmetry is, on the other hand, that
at generic vev the supersymmetric is spontaneously
broken; one finds some discrete choices of the holonomy
vev with the supersymmetry intact.
In either case, the process of the dimensional
reduction, as the circle size is taken to zero, is
typically ambiguous until we specify at which holonomy
vev this is done. When the holonomy is nontrivial, the
net effect is that of the Wilson line symmetry breaking.

When the space is compact or, more
precisely, has no more than two  extended directions, on the other hand,
the special nature of the holonomy variables manifest somewhat
differently, as they must be integrated over for the path
integral. For example, the localization for the twisted partition
functions produces integration over gauge holonomy variables
at the end of the procedure.
This means that one must be rather careful in taking
a small radius limit.  If one  naively replaces this integration
over the holonomy, living in the Cartan torus, by one
over ${\mathbb R}^\mathrm{rank}$, the Cartan
subalgebra, one ends up computing
partition function of a  dimensionally reduced theory
in one fewer dimension, with the vev of the holonomy
variable naively frozen at the identity.

As we commented already, however, dimensional reduction of a single
supersymmetric gauge theory on a circle may produce distinct
gauge theories in one fewer dimension, depending on what
holonomy vev's are available and chosen. For partition function
computations on a compact spacetime with a circle, then, this
ambiguity of the dimensional reduction must also manifest.
How does this happen?
Since the original integration range is over the Cartan
torus rather than the Cartan subalgebra and since the
periodic nature of the holonomy variables is not to be
ignored so easily, the answer  is quite
clear: As we scan the holonomy
along the Cartan torus, we often find special places
where the Wilson line symmetry breaking leads to
supersymmetric gauge theories in one fewer dimensions.

This translates to the supersymmetric partition function
$\Omega_d^G$ of theory $G$ in $d$-dimensions reducing,
in an appropriate scaling limit, to
a discrete sum of $(d-1)$-dimensional partition functions
${\cal Z}^H_{d-1}$ of theories $H$'s sitting at special
holonomies $u_H$, modulo some prefactors, as
\bea
\Omega_d^G \quad\rightarrow \quad \sum_{u_H} \sim {\cal Z}_{d-1}^H\ ,
\eea
where these $u_H$'s are distributed discretely
along the periodic Cartan torus. In the vanishing radius
limit, distinct $u_H$'s are infinitely far from one another,
so that taking the naive limit of replacing the holonomies
by scalars amounts to concentrating on a small
neighborhood near a single $u_H$. Since $u_H\neq 0$
would be infinitely far away from $u_H=0$ from the
perspective of dimensionally reduced theories, one is
often mislead to consider the theory at $u_H=0$, tantamount to
replacing the Cartan torus  by the Cartan subalgeba, and
ends up computing a wrong scaling limit of $\Omega_d^G$.

We will call these special holonomy values $u_H$'s
(and the supersymmetric theories sitting there) the holonomy saddles,
or $H$-saddles. Ref.~\cite{Hwang:2017nop} had
introduced this concept and thereby resolved a
fifteen-year-old puzzle\cite{Kac:1999av,Staudacher:2000gx,Pestun:2002rr}
on Witten indices of 1d
pure Yang-Mills theories \cite{Yi:1997eg,Sethi:1997pa,Green:1997tn,Moore:1998et,Lee:2016dbm,Lee:2017lfw};
in retrospect, the puzzle had originated from a simple misconception
that only the naive $u_H=0$ saddle (and its images under the
shift by the center) contributes to the right hand side.
Since the holonomy moduli space is present universally
for spacetimes with a circle, at least classically, and since the holonomy must
be integrated over for compact enough space, it is clear
that this $H$-saddle phenomenon will occur for twisted
partition functions regardless of spacetime dimensions.

For field theory Witten indices \cite{Witten:1981nf},
for example, $H$-saddles dictate how the Witten indices
of gauge theories in adjacent dimensions could be related.
Witten indices can easily differ in different dimensions despite the standard
rhetoric that compactification on torus does not change
the number of vacua. A well-known modern  example of
such disparities is how the 1d wall-crossing phenomena
does not manifest in 2d elliptic genera.
$H$-saddles now give us a rather concrete way to relate
such topologically protected quantities across dimensions,
in a very definite manner.

The importance of the holonomy in relating supersymmetric
theories between different dimensions has been noted
elsewhere, if somewhat sporadically.  Notable examples are
due to Aharony and collaborators \cite{Aharony:2013dha, Aharony:2017adm}
who observed how a Seiberg-dual pair of 4d/3d theories may
translate to multiple such in 3d/2d as well as an even earlier
work in Ref.~\cite{Aganagic:2001uw} where, again, a 2d
limit of a 3d mirror symmetry is explored.
Our study can be viewed as an effort to explore such
phenomena much more systematically and concretely,
now armed with varieties of exact partition functions,
and to consider other ramifications.
Also related are Refs.~\cite{Ardehali:2015bla,DiPietro:2016ond}
which found exceptions to the purported universal connection
between the Cardy exponents and the anomaly coefficients
\cite{DiPietro:2014bca}. What we find here is that such a
universal expression is often an artifact of ignoring
$H$-saddles other than the naive one at $u_H=0$ and that when
the theory comes with matter fields in gauge representations
bigger than the defining ones, this ``exception" tends to
occur generically for all acceptable spacetimes, including
$S^1\times S^3$. Furthermore, we will find similar
failures for the Casimir limit in general, although
this side proves to be more subtle.

We wish to emphasize that this phenomenon is inherent to
the supersymmetric gauge theories themselves, rather
than merely a property of the partition functions thereof.
Note that the latter quantities need compact spacetime for
their definition. When the spacetime has at least three
noncompact directions, these special values of the
holonomy give various superselection sectors where
the theories in one less dimensions are equipped with
supersymmetry intact at quantum level.
Nevertheless, the partition functions in general and
the Witten indices in particular offer handy tools for
classifying these special holonomies, which is why we
concentrate on computation of these quantities in this note.

This note is organized as follows. In the rest of this
introductory section, which also serves as a rough summary,
we will overview supersymmetric twisted partition functions
and give a broad characterization of $H$-saddle phenomena.
This phenomenon of $H$-saddles and their consequences
will be studied in the subsequent sections for a large class of
4d $\cN=1$ theories defined on compact spacetimes which are
$T^2$ fiber bundles over smooth Riemannian surfaces.

Section 2 will review a recent construction of A-twisted partition
functions in such backgrounds, and recall the detailed computational
procedure. This is then extended to the so-called ``physical"
backgrounds, one special case of which is the superconformal
index (SCI). Section 3 will classify the Bethe vacua in the small
and the large $\tau$ limits.
The Bethe vacua are easily seen to be clustered into subfamilies,
each of which can be regarded as the Bethe vacua of
some 3d theories sitting at special value of the holonomy.
Although the latter viewpoint is physically better motivated
in the small $\tau$ limit, which we can really view as
a compactification to 3d, the other limit of large $\tau$
follows the same pattern thanks to $SL(2,{\mathbb Z})$ property
of the fiber $T^2$. Even when the $SL(2,\mathbb Z)$ is not
available, such as in SCI's, such clustering of Bethe vacua
do occur as well, although, as we will see in Section 4.

These limiting behaviors of Bethe vacua imply
that a 4d gauge theory typically decomposes
into a disjoint sum of several, potentially distinct 3d theories:
The 3d limit of 4d supersymmetric partition functions becomes
 a sum of partition functions of these 3d theories, albeit with
extra exponential factors. A special case of this is the
4d Witten index, re-expressed as a sum of Witten indices of
the associated 3d theories at $H$-saddles, clearly without
the extra exponential prefactors. In Section 4, we explore
such limits for various background geometries and spacetime.
One noteworthy corollary here is that  the Cardy exponents,
and even the Casimir energies, to a lesser degree, would generally deviate
from the existing proposals \cite{DiPietro:2014bca,Bobev:2015kza,Closset:2017bse},
connected to various 4d anomalies.  As we will see these
proposals are often tied to the naive $u_H=0$ saddle
which may or may not be the dominant saddle. We should note,
however, that the Casimir limit of SCI's is somewhat special in that
the microscopic derivations in Refs.~\cite{Martelli:2015kuk,Assel:2015nca}
and the anomaly connection thereof proved to be robust, despite
the presence of nontrivial $H$-saddles. We comment on this toward
the very end of this note.

\subsection{Twisted Partition Functions and the Euclidean Time}

Twisted partition functions, to be denoted by $\Omega$
throughout this note, are obtained by computing
the partition function with an insertion of the chirality operator
$(-1)^{\cal F}$. A requisite for $(-1)^{\cal F}$ is that there is a
notion of natural Euclidean time coordinate, forming a circle $S^1$.
With the natural ${\mathbb Z}_2$ action of supersymmetry, say,
${\cal Q}$, which anticommutes with the chirality operator, this
insertion allows generic bosonic states cancel against
fermionic states, and leaves behind a special subset of the Hilbert space.

When the theory is suitably gapped and the space is taken to
be $T^{d-1}$, this quantity would compute the Witten index \cite{Witten:1981nf},
integral and enumerative of supersymmetric ground states.
In recent years, sweeping generalizations of such Index-like
quantities have been proposed with the accompanying computational
tricks under the banner of the localization. The superconformal
index \cite{Romelsberger:2005eg,Kinney:2005ej} is one such class of well-known and much-computed objects,
while the elliptic genera in 2d \cite{Benini:2013nda,Benini:2013xpa} and the refined Witten indices
in 1d have been developed to a very sophisticated level \cite{Hori:2014tda,Hwang:2014uwa}.

The length of Euclidean time circle, $\beta$, may be interpreted as
the inverse temperature. For the twisted version, however, this
parameter is often argued to disappear from the end result, since
supercharges ${\cal Q}$ act as a one-to-one map for positive
energy bosonic and fermionic states.
This disappearance is, of course, a desired feature
of the index, since the latter was designed, to begin with,
to count Bose-Fermi asymmetry of the ground state sector.
The twisted partition functions
\bea\label{twisted}
{\rm Tr}\;(-1)^{\cal F}e^{-\beta {\cal Q}^2}
\eea
are thus argued to be projected to the ground
state sector
\bea
{\rm Tr}_{{\rm kernel}({\cal Q})}\;(-1)^{\cal F}\ ,
\eea
which is necessarily integral and enumerative.

This is, however,
not quite true in general. If the theory admits continuum
spectrum whose energies are bounded below by $E_{\rm gap}>0$,
the trace (\ref{twisted}) actually produces
\bea\label{gap}
{\rm Tr}\;(-1)^{\cal F}e^{-\beta {\cal Q}^2} =
{\rm Tr}_{{\rm kernel}({\cal Q})}\;(-1)^{\cal F}+O(e^{-\beta E_{\rm gap}})\ .
\eea
This subtlety is relatively easy to handle since one
may be able to scale $E_{\rm gap}\rightarrow +\infty$
first, without affecting the ground state counting.
When $E_{\rm gap}=0$, on the other hand, separating out the continuum contributions
becomes something of an art. One popular scheme in the face of
such gapless asymptotic directions is to insert the chemical
potentials $\nu$'s for global symmetries $F$'s,
\bea
{\rm Tr}\;(-1)^{\cal F}e^{\,\nu F}e^{-\beta{\cal Q}^2}\ ,
\eea
where $[F,{\cal Q}]=0$ is needed for this quantity to remain
controllable. We may even have $F$ involving an $R$-charge,
as long as we choose one particular supercharge ${\cal Q}$ carefully
so that the two mutually commute. In many practical examples,
coming out of string theory, this option is available and exploited.

Although such an insertion of chemical potentials may appear
an innocent device to keep track of global charges of states,
this is true only for theories suitably gapped to begin with.
With gapless theories, this chemical potential modifies the
Lagrangian in such a way that asymptotic directions that transform
under $F$ become massive. Since one gaps the asymptotic
dynamics artificially, one should not expect
the twisted partition function to behave nicely in the $\nu =0$
limit.

Recovering information about the original theory prior
to turning on $\nu $ is hardly straightforward although
well-established routines exist for a handful classes of
theories. The Atiyah-Patodi-Singer
index theorem, applicable to non-linear sigma models onto
manifolds with boundary, is one such classic example while a more recent
such is the 1d gauged quiver quantum mechanics as explained
in Ref.~\cite{Lee:2016dbm}. Beyond these few, however, no general
prescription is known. Despite such difficulties, the
twisted partition functions of such mass-deformed theories
proved to be very useful for some tasks, e.g., most notably,
checking Strong-Weak dualities \cite{Intriligator:2013lca,Benini:2015noa,Benini:2016hjo,Closset:2016arn,Closset:2017zgf,Closset:2017bse}.

What do we do to actually
evaluate such objects? The popular trick of the localization
naturally enters the story when chemical potentials are turned on.
The chemical potentials $\nu $ tend to push the dynamics
to a small subset of the configuration space or even to a small
part of the spacetime; the localization method is then invoked
to amplify this effect maximally, whereby the path integral is
reduced to that of Gaussian path integral followed by finite number
of leftover zero-mode integrals from vector multiplets.

An interesting fact about the localization routine
is that, in the final expression, $\beta$ as in  $e^{-\beta{\cal Q}^2}$
automatically drops out. This may happen because
the system is fully gapped by $\nu $ so that the naive Bose-Fermi
cancelation works perfectly. In fact, this is the case for
main examples of this note, namely 4d
$\cN=1$ theories which are maximally mass-deformed by $\nu $'s.
As such, we will work with
\bea
\Omega(\nu )\;\;\equiv\;\;
{\rm Tr}\;(-1)^{\cal F}e^{\,\nu F}e^{-\beta{\cal Q}^2}\biggr\vert_{\rm localization}\ .
\eea
Sometimes this lift of the asymptotic flat direction by $\nu $
is incomplete, which tends to happen in odd spacetime dimensions.
In such cases, the localization still removes $\beta$ by computing,
implicitly, a limit of $\beta\rightarrow 0$,
\bea
{\rm Tr}\;(-1)^{\cal F}e^{\,\nu F}e^{-\beta{\cal Q}^2}\biggr\vert_{\rm localization}=\;\;
\lim_{\beta\rightarrow 0} {\rm Tr}\;(-1)^{\cal F}e^{\,\nu F}e^{-\beta{\cal Q}^2}\ .
\eea
This has been first noted for 1d systems \cite{Lee:2016dbm}
and further checked in Ref.~\cite{Hwang:2017nop}.

Although we have described how the Euclidean time span $\beta$
naturally drops out in the localization
computation, the resulting twisted partition function $\Omega$
can actually retain $\beta$ indirectly, via the chemical
potential $\nu $  understood as a holonomy associated
with an external flavor gauge field,
\bea
i\partial_t \quad\rightarrow\quad i\partial_t + \frac{\nu  F}{\beta}\ .
\eea
In the small $\beta$ limit, one has an option of keeping
$\nu $ finite or keeping the alternate variable $\tilde\nu $ finite with
\bea
\nu =\beta\, \tilde\nu\ .
\eea
The so-called ``real" masses in 3d are, for example, nothing but such finite $\tilde\nu $.

We can think about something similar for
the gauge holonomy variables, $u$,  which enter the
localization formulae for $\Omega$ as
\bea\label{d}
\Omega^G(\nu )\;\; =\;\; \int d^\mathrm{rank}u\; g_G(u;\nu )\ .
\eea
We introduced the label $G$ to denote the theory and
$g_G(u,\nu )$ is from the Gaussian integrals over non-zero-modes.
If we introduce the similarly rescaled variables $\tilde u=u/\beta$
in the small $\beta$ limit, the following object
where the integral is taken over $\tilde u$ instead of $u$,
\bea
\sim\;\; \int d^\mathrm{rank}\tilde u\; \lim_{\beta\rightarrow 0}\;\beta^{\mathrm{rank}}g_G(\beta \tilde u;\beta \tilde\nu )
\eea
appears naturally. Since $\beta$ is taken to be arbitrarily
small, the periodic nature of $u$ is now lost. What would such an
integral compute?

To be precise,  let us consider a spacetime of type
$S^1\times { M}_{d-1}$. In the small radius limit,
the dimensional reduction on $S^1$ produces a theory
on ${M}_{d-1}$ with the same
field content as the original theory. We will label this
theory on ${M}_{d-1}$ by the same label $G$,
whose partition function would also produce a localized
path integral as
\bea\label{d-1}
{\cal Z}^G(\tilde\nu )\;\;=\;\; \int d^\mathrm{rank}\tilde u\;f_G(\tilde u;\tilde\nu )\ .
\eea
Past experiences with such objects tell us that the limit
is often equipped with extra exponential factor,
\bea
\beta^{\mathrm{rank}} g_G(\beta \tilde u;\beta \tilde\nu )
\quad\rightarrow\quad e^{S^{\rm Cardy}_G/\beta+{\rm subleading\; terms}} f_G(\tilde u;\tilde\nu )
\qquad {\rm as} \;\; \beta\rightarrow 0 \ ,
\eea
where $S^{\rm Cardy}_G$ is the Cardy exponent \cite{Cardy:1986ie}.
Then, the naive expectation is
\bea\label{naive}
\Omega^G(\beta \tilde\nu ) \quad\rightarrow\quad
e^{S^{\rm Cardy}_G/\beta+{\rm subleading\; terms} }\times
{\cal Z}^G(\tilde\nu )\qquad {\rm as}\;\; \beta\rightarrow 0\ ,
\eea
where the two partition functions were computed
for the one and the same gauge theory, $G$, only
in two different dimensions. The exponent $S^{\rm Cardy}_G$
would dictate ``high temperature behavior" of the
twisted partition function.

\subsection{Holonomy Saddles: An Overview }

However, comparing (\ref{d}) and (\ref{d-1}), one easily realizes
that this is too rash. A limiting formula like (\ref{naive})
would hold if and only if the toroidal $du$ integration in
(\ref{d}) can be opened up to a planar integration in (\ref{d-1});
since this is a discontinuous process, this may be justified only
if, in the small $\beta$ limit, the infinitesimal region around
$u=0$ contributes dominantly to the integral.

As a simple counterexample, which may look trivial but is
illuminating nevertheless, consider an $SU(2)$ gauge
theory with matter multiplets with integral isospins
only. Suppose that we choose the range of $u$ suitable for
the odd isospins, say $[0,1)$ in our convention where
weight vectors are normalized to be integral and
the holonomies are divided by $2\pi$.
The integrand $g(u;z)$ would then be invariant under the shift
related by the center, $u\rightarrow u+1/2$;
the expansion of $g$ around $u=1/2 $ will look exactly the same
as that around $u=0$, so that the integral near $u=0$ and
that near $u=1/2$ contribute exactly the same amount.
Although this particular problem is easily
countermanded by an overall factor 2, it does warn us of a
generic danger in confining ourselves to small regions near $u=0$
when $u$ is a periodic variable.

What happens generically is that the small $\beta$ limit of
$\Omega(\beta \tilde\nu)$ is actually a sum of ${\cal Z}$'s for
several disjoint theories on ${M}_{d-1}$, such that the limit
has the form
\bea\label{H}
\Omega^G(\beta \tilde\nu) \quad\rightarrow\quad \sum_{u_H}
e^{S^{\rm Cardy}_H/\beta+{\rm subleading\; terms}}
\times {\cal Z}^H(\tilde\nu)\qquad {\rm as}\;\; \beta\rightarrow 0\ ,
\eea
instead of (\ref{naive}). The summand is labeled by special
holonomies values $u_H$ around which the dimensional
reduction gives a theory $H$, with potentially smaller field
content than the naive dimensional reduction of the original theory $G$.
The integration over the toroidal $u$'s reduces to
patches of planar integrations near such $u_H$'s
while contributions from the rest become suppressed by
$e^{-1/\beta}$. The accompanying limit in
the localization formulae should be similarly
\bea\label{d-1:H}
&&{\beta^{\mathrm{rank}}}g_G(u_H+\beta \tilde u;\beta \tilde\nu)
\;\;\rightarrow\;\; e^{S^{\rm Cardy}_H/\beta+{\rm subleading\; terms}}
\times f_H(\tilde u;\tilde\nu)\qquad {\rm as} \;\; \beta\rightarrow 0 \cr\cr
&&{\cal Z}^H(\tilde\nu)\;\;=\;\; \int d^\mathrm{rank}\tilde u\;f_H(\tilde u;\tilde\nu)\ .
\eea
The discrete locations $u_H$ are infinitely separated from
one another, in the limit of $\beta\rightarrow 0$,  and
thus cannot be captured by the $\tilde u$ integration near the
origin alone.

With at least one nontrivial $u_H$, one must ask which
of these saddles are dominant in the small $\beta $ limit;
one might have expected that the theory $G$ at the naive saddle at $u_H=0$
is the dominant one, given its largest light field content,
but it turns out this is generally false. In particular,
when ${M}_{d-1}$ is $T^{d-1}$ whereby $\Omega^G$ and ${\cal Z}^H$ would
both compute the  Witten indices, each of admissible $u_H$
generically contributes on equal footing. In other words,
(\ref{H}) would reduce to
\bea\label{HIndex}
{\cal I}_{d}^G(\beta \tilde\nu) \;\;\rightarrow \;\;
 \sum_{u_H}\;{\cal I}_{d-1}^H(\tilde\nu)\qquad {\rm as} \;\; \beta\rightarrow 0\ .
\eea
This means that $u_H$ must be such that the theory $H$
there must have supersymmetric vacua.\footnote{When
the theory possesses gapless asymptotic sector, so
that the twisted partitions do not compute the true
index, this condition should be relaxed since the
twisted partition functions capture the so-called
bulk part of the true index.}

Let us take $d=4$ $\cN=1$ theories, which will be
our main examples.
With generic holonomies, $u$, the 3d theory would be
a product of free $U(1)$'s whose vacuum manifolds are
generically lifted by combination of induced Fayet-Iliopoulos (FI) constants or
Chern-Simon levels. Light charged multiplets, say with the
charge $\lambda$ with respect to the Cartan $U(1)$'s,
would be needed for vacua with unbroken supersymmetry.
This constrains the position $u$ to quantized
values, $u_H$,
\bea
\lambda\cdot u_H \quad\in\quad {\mathbb Z}
\eea
for each such $\lambda$. In this manner, a contributing
$H$-saddle is equipped with a set of unbroken charges
$\lambda$'s, which in turn defines the 3d theory $H$
at $u_H$, modulo UV couplings in the 3d sense
inherited  and computable from the original 4d theory $G$. What we
described here is a little simplified; it turns out
that when the matter content is not symmetric under
charge conjugation, one can actually have an $H$-saddle
with decoupled $U(1)$'s or pure Yang-Mills sectors,
as long as appropriate Chern-Simons coefficients are
generated from integrating out heavy modes.
What remains unchanged,  though, is that contributing
$u_H$'s occur discretely. See Section \ref{saddles}
for a complete characterization of $H$-saddles.

What we described above is a generic feature of
gauge theories, due to the special roles played by
the holonomy variables: The toroidal nature of  the
holonomy variables appears lost in the small radius
limit, yet the periodic nature should not be ignored.
Integrating over such holonomy variables, such as
for gauge theories on compact spacetime with a circle,
we must remember to keep a careful track of these
holonomies. For supersymmetric partition functions,
it so happens that there are multiple saddles which
contribute to the total expression, each of which can
be understood as a partition function of some other
theories in one fewer dimensions.

In this note, we will  consider implications
of $H$-saddles in the context of 4d $\cN=1$ theories
on $T^2$ fibred over Riemannian surfaces of arbitrary genus $\Sigma_g$.
General partition functions of this class were given very
recently via the so-called Bethe Ansatz Equation (BAE) \cite{Closset:2017bse}.
In this approach one first consider compactification
on $T^2$ reducing the system to 2d, and vacua and
partition functions are found via the effective 2d
twisted superpotential of Coulombic variables. The
vacua thus found is called Bethe vacua \cite{Nekrasov:2009uh}.
As such, this construction works for a restricted class
of 4d theories, where, given the matter content,
the superpotential is appropriately suppressed
to allow maximal flavor symmetry.
On the other hand, the construction
is ideal for the investigation of the $H$-saddle phenomena since
the latter turns out to be quite manifest in the classification
of BAE vacuum solution. $T^2$ fiber has two circles,
whose relative size is encoded in the complex structure
$\tau$. In the large and the small $\tau$ limit, one of
the two circles becomes very small relative to the
other, and as such the phenomenon of $H$-saddle emerges on
the smaller of the two circle directions.

The 2d twisted superpotential ${\cal W}$ is
naturally a function of the pair of the holonomies
along $T^2$, packaged into rank-many complex coordinates $u$.
Bethe vacua are particular holonomy values, $u_*$, where
$e^{2\pi i\partial{\cal W}}=1$. This vacuum condition
is periodic under integral shifts, $u\rightarrow u+n+m\tau$
where $\tau$ is the complex structure of $T^2$, which is
a gauge equivalence. What we will discover is that these
Bethe vacua appear in clusters, scattered at discrete
places in the unit cell,
\bea
u_*/\tau \simeq \tilde u_H+ \tilde \sigma_*\ , \qquad u_*\simeq u_H+ \sigma_*\ ,
\eea
where $\tilde u_H\sim 1/\tau$ and $u_H \sim \tau$ with coefficients
between 0 and 1 for $\tau\rightarrow i0^+,\;i\infty$ limits,
respectively. Each such  $H$-saddle would come with multiple
and finite $\tilde\sigma_*$'s and  $\sigma_*$'s, which
represents supersymmetric vacua in the reduced 3d theory at
such $H$-saddles. Depending on which circle is called the
Euclidean time, the limit will also compute the Cardy exponents
or the Casimir energies at each of such $H$-saddles. The
question of which saddle dominates becomes a nontrivial
issue, generically compromising folklore on universality
of such exponents. We will revisit the asymptotics of
the partition functions in Section 4.

\section{4d ${\cal N}=1$ Partition Functions and BAE}
\label{sec:review}

Recently the supersymmetric partition function of a
four-dimensional $\mathcal N = 1$ theory on $\mathcal M_4$
was discussed \cite{Closset:2017bse} where $\mathcal M_4$ is a torus bundle
over a Riemannian surface $\Sigma_g$:
\begin{align}
T^2 \;\;\rightarrow\;\; \mathcal M_4\;\; \rightarrow\;\; \Sigma_g\ .
\end{align}
The partition function is obtained by considering an
A-twisted theory on $\Sigma_g$ via nontrivial background
flux $\mathfrak n_R = g-1$ for the  $U(1)_R$ symmetry group.
Here we give a quick summary of results in Ref.~\cite{Closset:2017bse}.
We will use $G$ to denote the $\cN=1$ theory itself,
while $\cG$ and $\mathfrak G$ are the gauge group and the
associated Lie Algebra, respectively.

Before we plunge into details, a cautionary remark is
in order. Much of what follows will be phrased in terms
of gauge holonomies on $T^2$, valued in two copies of
the Cartan torus.  As was emphasized by Witten and
others
\cite{Witten:1997bs,Keurentjes:1999mv,Keurentjes:1999qf,Kac:1999gw,Witten:2000nv}, however, the space
of connections on $T^2$ and higher dimensional torii
can in general admit disconnected components, even when
the gauge group is connected.
Such possibilities are not taken into account in the
computations outlined below, so we will
confine ourselves, in this note, to theories with $\cG=SU(r+1), Sp(r)$.

\subsection{A-Twisted Background}

Compactifying on $T^2$, one has a two-dimensional
$\mathcal N = (2,2)$ supersymmetric theory with infinite
Kaluza-Klein modes.
Performing the path integral via localization, summing
over the magnetic flux sector, and then evaluating the
resulting residue formulae, the partition function is
written universally as a sum over the so-called Bethe
vacua,
\bea
\label{eq:pf}
 \Omega= \sum_{u_* \in \mathcal S_\text{BE}} \mathcal F_1 (u_*,\nu;\tau)^{p_1} \, \mathcal F_2 (u_*,\nu;\tau)^{p_2} \, \mathcal H(u_*,\nu;\tau)^{g-1} \, \prod_\alpha \Pi_A(u_*,\nu;\tau)^{\mathfrak n_\alpha} \ .
\eea
where $p_1, \, p_2 \in \mathbb Z$ are the two
Chern numbers of the circle bundles:
\begin{align}
p_1 = \frac{1}{2 \pi} \int_{\Sigma_g} dA_{KK_1}, \qquad p_2 = \frac{1}{2 \pi} \int_{\Sigma_g} dA_{KK_2} \ .
\end{align}
$\tau = \tau_1+i \tau_2$ is the modular parameter of the torus with
$\tau_2 = \frac{\beta_2}{\beta_1}$, where $\beta_1$ and $\beta_2$ are two radii
of the torus. Note that one can perform a modular transformation of $T^2$ such that
$(p_1,p_2) = (p,0)$.

Now let us see how field theory data enter this universal formula.
The basic  variables are holonomies along $T^2$.
The gauge holonomies $u_1, u_2$, along these two circles
\bea
a_1 = \frac{1}{2 \pi} \int_{S^1_{\beta_1}} A\ , \qquad a_2 = \frac{1}{2 \pi} \int_{S^1_{\beta_2}} A\ ,
\eea
of $2\pi$ period each, are combined to
\bea
u=a_1\tau - a_2
\eea
and, similarly for the complexified flavor holonomy
\begin{align}
\nu = a_1^{(F)}\tau-a_2^{(F)}\ .
\end{align}
All of these holonomy variables obey
\begin{align}
u_a \sim u_a+1 \sim u_a+\tau, \qquad \nu_A \sim \nu_A+1 \sim \nu_A+\tau\ ,
\end{align}
under the respective large gauge transformations.

The effective action of the theory is fully
governed by two holomorphic functions $\mathcal W$ and $\Omega$, which are called
the effective twisted superpotential and the effective dilaton. Each component in
\eqref{eq:pf} and \eqref{eq:BE} is then obtained from those two quantities,
\bea
\mathcal F_1(u,\nu;\tau) &=& \exp\left[2 \pi i \frac{\partial \mathcal W}{\partial \tau}\right]\ , \\
\mathcal F_2(u,\nu;\tau) &=& \exp\left[2 \pi i \left(\mathcal W-u_a \frac{\partial \mathcal W}{\partial u_a}-\nu_A \frac{\partial \mathcal W}{\partial \nu_A}-\tau \frac{\partial \mathcal W}{\partial \tau}\right)\right]\ , \\
\mathcal H(u,\nu;\tau) &=& e^{2 \pi i \Omega(u,\nu;\tau)} \left(\det_{ab} \frac{\partial^2 \mathcal W(u,\nu;\tau)}{\partial u_a \partial u_b}\right)\ , \\
\Phi_a(u,\nu;\tau) &=& \exp\left[2 \pi i \frac{\partial \mathcal W}{\partial u_a}\right]\ , \qquad \Pi_A(u,\nu;\tau) = \exp\left[2 \pi i \frac{\partial \mathcal W}{\partial \nu_A}\right]\ .
\eea
For semi-simple $\cG$, the W-bosons and their superpartners do not contribute to the effective twisted superpotential, and only charged chiral multiplets contribute. The contribution of a single chiral multiplet is given by
\begin{align}
\label{eq:W}
\mathcal W_\Phi = -\frac{u^3}{6 \tau}+\frac{u^2}{4}-\frac{u \tau}{12}+\frac{1}{24}+\frac{1}{(2 \pi i)^2} \sum_{k = 0}^\infty \left(\mathrm{Li}_2 \left(x q^k\right)-\mathrm{Li}_2 \left(x^{-1} q^{k+1}\right)\right)\ ,
\end{align}
where $\mathrm{Li}_2$ is a polylogarithm function.
We have defined $x = e^{2 \pi i u}$ and $q = e^{2 \pi i \tau}$.

Classification of the Bethe vacua $\mathcal S_\text{BE}$ starts
with solving
\bea
1=\Phi_a\ ,
\eea
which can be expressed more explicitly as,
\bea
&&\Phi_a (u,\nu;\tau) = \prod_i\prod_{\rho_i}
\Psi(\rho_i\cdot u+\nu_i;\tau)^{\rho^a_i}\ ,\cr\cr
&&\Psi(w;\tau)\equiv e^{-\pi i w^2/\tau}\theta(w,\tau)^{-1}\ ,
\cr\cr
&&\theta(w,\tau) = q^{1/12}t^{-1/2}\prod_{k\ge 0}(1-tq^k)(1-t^{-1}q^{k+1})\ ,
\eea
with $t=e^{2\pi i w}$, where $\nu_i\equiv \nu\cdot F_i$
is the net sum of flavor chemical potentials for the $i$-th multiplet with flavor charges $F_i$.
The product is over chiral multiplets, labeled
by $i$ for each gauge multiplet and the charges $\rho_i$ thereof
with respect to the Cartan.
Then,  $S_\text{BE}$, which is nothing but the set
of the supersymmetric vacua for the 2d twisted superpotential,
may be defined as
\begin{align}
\label{eq:BE}
\mathcal S_\text{BE} = \left\{u_*\;\vert\;\Phi_a(u_*,\nu;\tau) = 1, \forall a, \quad w \cdot u_* \neq u_*, \forall w \in W_\cG\right\}/W_\cG\ .
\end{align}
The additional constraint, that vacua invariant under
any part of the Weyl group $W_\cG$ are to be ignored,
has been noted in the past literature, most notably
Refs.~\cite{Hori:2000kt,Hori:2006dk}.

With the explicit form of $\mathcal W$ in \eqref{eq:W},
one can similarly compute the rest. The effective dilaton contribution is given by
\begin{align}
\label{eq:dilaton}
e^{2 \pi i \Omega} = \left(\prod_i \prod_{\rho_i \in \mathfrak R_i} \Psi (\rho_i \cdot u+\nu_i;\tau)^{r_i-1}\right) \left(\eta(\tau)^{-2 \, \mathrm{rank}(\cG)} \prod_{\alpha \in \mathfrak g} \Psi (\alpha \cdot u;\tau)\right)\ ,
\end{align}
where the product in the second parentheses is taken over the roots of $\mathfrak g = \mathrm{Lie}(G)$. From the effective dilaton \eqref{eq:dilaton} and the effective twisted superpotential \eqref{eq:W}, one can obtain the handle-gluing operator $\mathcal H$ as well.
The explicit forms of the fibering operators $\mathcal F$ and the flavor
flux operators $\Pi_{\alpha}$ are
\begin{align}\label{F's}
\mathcal F_{1,2} (u,\nu;\tau) &= \prod_i \prod_{\rho_i \in \mathfrak R_i}
 \Xi_{1,2} (\rho_i \cdot u+\nu_i;\tau)\ , \\
\Pi_A (u,\nu;\tau) &= \prod_i \prod_{\rho_i \in \mathfrak R_i}
\Psi (\rho_i \cdot u+\nu_i;\tau)^{\omega_i^\alpha}\ ,
\end{align}
where we have defined
\begin{align}\label{Cascade}
\Xi_1 (w;\tau) &= e^{\frac{\pi i}{3 \tau^2} w^3-\frac{\pi i}{6} w} \, \Gamma_0(u;\tau)\ , \\
\Xi_2 (w;\tau) &= e^{2 \pi i \left(\frac{w^3}{6 \tau}-\frac{w^2}{4}+\frac{w \tau}{12}+\frac{1}{24}\right)} \, \prod_{k = 0}^\infty \frac{f(w+k \tau)}{f(-w+(k+1) \tau)}\ ,
\end{align}
with
\begin{gather}\label{Gamma0}
\Gamma_0(w;\tau) = \prod_{n = 0}^\infty \left(\frac{1-t^{-1} q^{n+1}}{1-t q^{n+1}}\right)^{n+1}, \\
f(w) = \exp \left[\frac{1}{2 \pi i} \mathrm{Li}_2 (e^{2 \pi i w})+w \log (1-e^{2 \pi i w})\right] \ .
\end{gather}
For the flux operator, the products are taken over every chiral multiplet and the weights of its representation. $\rho_i^a$ is the $a$-th Cartan charge of the gauge weight $\rho_i$ while $\omega_i^\alpha$ is the $\alpha$-th Cartan charge of the flavor weight $\omega_i$. With nontrivial background flux $\mathfrak n_\alpha$ for the flavor group, the flavor flux operator $\Pi_A$ contributes to the partition function as in \eqref{eq:pf}.

Before proceeding, however, we should mention a few caveats. The most
obvious is the presence of non-anomalous $U(1)_R$ symmetry.
Since this symmetry is used for the topological A-twisting \cite{Witten:1988xj},
one must further require the $U(1)_R$ charges of chiral
multiplets $r_i$ be integral. As such, neither for pure $\cN=1$
Yang-Mills theories nor for typical $\cN=1$ superconformal
theories would this methodology be applicable. Later, however, we will go
to a slightly different class of spacetime, with the same
geometry but different fluxes, so that the
integrality of the $U(1)_R$ charges can be relaxed.
There, the partition function formula should be
applicable to $\cN=1$ superconformal theories with
$a$-maximized $R$-charges.

A less obvious caveat, although quite rampant in the
exact partition function computations, comes from the flavor
chemical potentials. As mentioned earlier, the latter means
that the theory is artificially mass-deformed, and that
we may not be able to recover physics of the original
undeformed theory easily. This danger is present in all
exact twisted partition function computations via the
localization, but perhaps a little more so in this class
since this computation turns on a chemical potential
for each and every chiral multiplet:
One must always take extreme care in interpreting
the results.

With $SU(N)$ theories with
$N_f$ fundamental flavors, for example, this current
computation would give simple numerical Witten index for
all $N_f$ if we take ${\cal M}_4=T^4$. However, such
theories are often equipped with a manifold of the vacuum
moduli, which, since the number of spacetime dimensions
is larger than two,  should have made the notion of the Witten
index ill-defined. One must really regard these partition
functions as probing theories that are compactified  on
$T^2$  with flavor holonomies necessarily turned on
along the two circles.

\subsection{Alternate Backgrounds and Superconformal Index}

So far we have considered an A-twisted theory, whose supersymmetric
background includes the nontrivial $U(1)_R$ gauge field of
\begin{align}
\nu_R = 0, \qquad \mathfrak n_R = g-1 \ .
\end{align}
On the other hand, there is another class of supersymmetric backgrounds
without the $U(1)_R$ flux, which is called ``physical gauge" in Ref.~\cite{Closset:2017bse}:
\begin{align}
\nu_R = \frac{1-g}{p} \tau, \qquad \mathfrak n_R = 0\ ,
\end{align}
with $g-1 \equiv 0 \mod p$.
This is achieved by starting with $p_1=p$ and $p_2=0$, and
taking a large gauge transformation on $R$-symmetry that
removes the $R$-flux in favor of the $R$-chemical potential.

Ref.~\cite{Closset:2017bse} proposed that the partition function in this
background can be written in a similar manner but
with different operators ${\cal H}$, ${\cal F}$, etc, which
will be our working assumption, below. The flux operators are
\bea
\Phi_a^\text{phys} (u,\nu,\nu_R;\tau) &= &\prod_i \prod_{\rho_i \in \mathfrak R_i} \Psi (\rho_i \cdot u+\nu_i+\nu_R (r_i-1);\tau)^{\rho_i^a}\ , \cr\cr
\Pi_A^\text{phys} (u,\nu,\nu_R;\tau) &= &\prod_i \prod_{\rho_i \in \mathfrak R_i} \Psi (\rho_i \cdot u+\nu_i+\nu_R (r_i-1);\tau)^{\omega_i^\alpha}\ ,
\eea
while the fibering operator for the circle 1 is
\bea
\mathcal F^\text{phys}(u,\nu,\nu_R;\tau) &= &\left(\prod_i \prod_{\rho_i \in \mathfrak R_i} \Xi_1(\rho_i \cdot u+\nu_i+\nu_R (r_i-1);\tau)\right) \\ \cr
&& \times\left((-1)^{\frac{l_R (l_R+1)}{2} \mathrm{rank}(\cG)} \eta(\tau)^{2 l_R \mathrm{rank}(\cG)} \prod_{\alpha \in \mathfrak G} \Xi_1(\alpha \cdot u+\nu_R;\tau)\right)\ ,\nonumber
\eea
where
\begin{align}
\nu_R = l_R \, \tau, \qquad l_R = \frac{1-g}{p} \in \mathbb Z\ .
\end{align}
In addition, in the physical gauge, there is no effective
dilaton contribution because $\mathfrak n_R = 0$. Thus, the
handle-gluing operator is simply the Hessian determinant,
\begin{align}
H^\text{phys} (u,\nu,\nu_R;\tau) &= \det_{ab}
\left(\frac{1}{2 \pi i} \frac{\partial
\log \Phi_a^\text{phys}}{\partial u_b}\right)\ .
\end{align}
As a result, the partition function in the physical gauge is given by
\begin{align}
\Omega^\text{phys} = \sum_{u_* \in \mathcal
S_\text{BE}^\text{phys}} \mathcal F_1^\text{phys}
(u_*,\nu,\nu_R;\tau)^{p} H(u_*,\nu,\nu_R;\tau)^{g-1}
\, \prod_\alpha \Pi_A^\text{phys} (u_*,\nu,\nu_R;\tau)^{\mathfrak n_\alpha}\ ,
\end{align}
where we restricted ourselves to the case $p_1=p$ and $p_2=0$,
with the Bethe vacua
\begin{align}
\label{eq:BE}
\mathcal S_\text{BE}^\text{phys} = \left\{u_*\;\vert\;
\Phi_a^\text{phys} (u_*,\nu,\nu_R;\tau) = 1, \forall a,
\quad w \cdot  u_* \neq u_*, \forall w \in W_\cG\right\}/W_\cG\ .
\end{align}
Furthermore, Closset et. al. advocated that
once we arrive at this so-called "physical gauge," the
integral restriction on $r_i$ can be lifted.

As such, this partition function is supposed to compute
a limit of the superconformal index \cite{Romelsberger:2005eg,Kinney:2005ej} if we take $g=0$ and $p=1$,
\bea
{\Omega}_{S^1\times S^3}(q; x)\equiv {\rm Tr}_{ S^3}
\left[(-1)^{\cal F} q^{2J+R} x^{G_F} e^{-\beta_2 H}\right]\ ,
\eea
where the pair of rotational chemical potentials that
enter the
usual superconformal index are identified. For the superconformal
index, the large and the small radius limits are already discussed in
the literature quite extensively \cite{DiPietro:2014bca,Assel:2015nca,Ardehali:2015bla,DiPietro:2016ond,Bobev:2015kza,Martelli:2015kuk}. We reexamine these
limits of the partition function using the Bethe formalism.
The subtlety with the holonomy should be again present, and
many results of the previous section carry over to the new
background verbatim.
Ref.~\cite{Closset:2017bse} initially motivated this construction
via a large $U(1)_R$ transformation, as outlined above,
from the A-twisted cases with $p\neq 0$. This cannot be really
considered a derivation since the non-integral values of $r_i$'s,
inevitable for $\cN=1$ superconformal field theories,
would be detrimental to such a process. On the other hand,
an alternate justification was also given by the same authors,
where these BAE expressions for $g=0, p=1$ is
transformed to the conventional form of the superconformal index
via contour manipulations. See Eq.~(\ref{unitcircle}).

\section{Bethe Vacua for Elongated $T^2$ }\label{saddles}

In this note, for the sake of convenience, we will regard
circle 2 the Euclidean time. Then the
large and the small Euclidean times correspond to,
respectively,
\bea
\tau = \cdots +i\frac{\beta_2}{\beta_1} \quad\rightarrow\quad i\infty \;{\rm or}\;\; i0^+\ .
\eea
The small $\tau$ limit can be viewed as compactification along 
circle 2, while the large $\tau$ limit would be viewed as the
compactification along circle 1. This interpretation is
possible as the size of the base $\Sigma_g$ appears nowhere in the
partition functions, and also because only the ratio of the two radii
appears. For either compactification, the Kaluza-Klein towers
will acquire a large mass shift at typical values of
holonomy. This means that the low energy effective theory in the
remaining three dimensions would be rank-many free $U(1)$
theories. At such a generic point we will find that the 3d
theory has supersymmetry spontaneously broken and
thus cannot contribute solutions to the BAE.

What we will find is that solving 4d BAE will produce vacua
clustered at  some discrete and special choices of the holonomy.
As $\tau\rightarrow i\infty$, these special holonomies $u_H$
line up along circle 1, while for the other limit
$\tau\rightarrow i0^+$ they line up along the circle 2.
At such special places $u_H$, i.e. at $H$-saddles, with the holonomy,
the gauge charge set of the chiral multiplet will split,
\bea\label{4dto3d}
\{\rho\}\;\;=\;\;\{\lambda\} \;\cup\; \{\hat\rho\}\ ,
\eea
where those chirals associated with $\lambda$'s will produce
light 3d chiral multiplets at the bottom of the KK tower, while 
those associated with $\hat\rho$ will produce KK towers
with no such light 3d field. The vector multiplets in the adjoint
representation would be also similarly decomposed, and a spontaneous
symmetry breaking by Wilson line will occur
\bea
\cG\quad\rightarrow\quad \cH\ ,
\eea
where the unbroken, 3d gauge group $\cH$ has the same rank
as the 4d gauge group $\cG$. The new 3d gauge theory $H$,
typically with smaller light field content, both vectors
and chirals, than the naive dimensional reduction of the
theory $G$, appears. The holonomy $u_H$'s and the new
theories there, we will collectively call $H$-saddles \cite{Hwang:2017nop}.

This classification of $H$-saddle, to be explained
in detail below, could include some special
cases. The case with $\{\lambda\}=\{\rho\}$,
for example at $u_H=0$,  would produce the 3d
theory with the same field content as the naive dimensional
reduction. Most of the related literature have assumed, effectively,
that this type of $H$-saddle is either the only kind or the
dominant one. The other extreme $\{\lambda\}=\emptyset$, or
more generally the cases where $\lambda$ do not span
the charge vector space, would produce 3d theory with
a pure gauge sector. Also, $H$-saddles that include Abelian
subgroup in $\cH$ require more care, since large 3d
FI constant can be generated even though the 4d theory had no
such Abelian factor. For the latter types of $H$-saddles,
a little more care must be given, which we will go through
in subsections 3.3. and 3.4.

While most of this section is devoted to the A-twisted case,
the ``physical" version is really no different. The BAE equations
remain identical to those of the A-twisted case, except the
additional $U(1)_R$ chemical potential $(r_i-1)l_R\tau$ for
the chiral fields. As such, the large $\tau$ limit of the
``physical" version requires extra care, which will be
addressed in Subsection 3.5.

A comment on a notation is in order, to avoid confusion.
For a holonomy variable $w$ with the natural periods,
$\tau$ and $1$, we will define its ``fractional" part,
$\{w/\tau\}$, as
\bea
{}\{w/\tau{}\}\equiv w/\tau-m\ ,
\eea
where
\bea\label{floor}
m\equiv \lfloor w/\tau\rfloor
\eea
is an integer
such that the real part of $w/\tau-m$ lies in $[0,1)$.
It follows that, for example,
\bea
{}\{\lambda\cdot u_H /\tau{}\}=0\ ,\qquad {}\{\hat\rho\cdot u_H /\tau{}\}\neq 0\ ,
\eea
in the large radius limit of A-twist cases
(\ref{4dto3d}), and
\bea
{}\{\lambda\cdot \tilde u_H /\tilde \tau{}\}=0\ ,\qquad {}\{\hat\rho\cdot \tilde u_H /\tilde \tau{}\}\neq 0\ ,
\eea
with $\tilde u\equiv u/\tau$ and $\tilde\tau=-1/\tau$,
in the small radius limit of A-twist cases.
Note that we also use the same curly bracket $\{\cdots\}$
as a symbol for sets, as is customary. Hopefully, the
distinction between these two is self-evident.

\subsection{$H$-Saddles in the Small $\tilde\tau=-1/\tau$ Limit }

We will start with $\tau\rightarrow i\infty$, or $q\rightarrow 0$,
although the discussion below may appear more natural in the
other limit of $\tau\rightarrow i0^+$. In the small $\tau$ limit, the role
of the two circles will be exchanged, so that our finding here will
carry over almost verbatim, via an $SL(2,{\mathbb Z})$ action.

We start by noting that there are two different types of
solutions to the 4d BAE equation in such a limit.
The first type comes from assuming $u$ remains finite
under the scaling, and as such we find
\bea\label{vanilla}
\Psi(w;\tau) & \rightarrow &\frac{ q^{-1/12}}{t^{-1/2}-t^{1/2}} \\\cr
\Phi_a (u;\tau)& \rightarrow & \Phi_a^{3d}\equiv \Lambda_G^a\prod_i\prod_{\rho_i}
\left[x^{-\rho_i/2}y_i^{-1/2}-x^{\rho_i/2}y_i^{1/2}\right]^{-\rho_i^a}\nonumber
\eea
with
\bea
\Lambda_G^a = q^{-\sum_{i,\rho_i} \rho_i^a/12}\ ,
\eea
where $y_i\equiv e^{2\pi  i\nu_i}$ and $x^{\rho_i}\equiv \prod_a x_a^{\rho_i^a}$.
The resulting BAE equations, $\Phi^{3d}_a=1$, look exactly the same as
the 3d BAE equations \cite{Benini:2015noa,Benini:2016hjo,Closset:2016arn,Closset:2017zgf} of a dimensionally reduced theory with
the same field content. The only unexpected feature is that
$2\pi i \tau$ now plays the role of a UV FI constant for the
trace-part $U(1)$ gauge field when the latter is present.

If $\sum \rho_i^a\neq 0$, the locations of the
solutions, $x_a$, could be too far away and conflict with the
above truncation. Thankfully, however, this never really happens
for 4d theories with no $U(1)$ factor, as is necessary for the
asymptotic freedom. With the gauge group $\cG$ being at most
a product of semi-simple Lie groups, we find
\bea
\Lambda_G^a=1
\eea
generally.\footnote{
This can be seen from the Weyl character formulae, with
the Cartan generators ${\cal C}$'s in an irreducible
representation $\mathfrak R$ of a semi-simple gauge group $\cG$,
\bea
\chi_R(e^{u})\equiv {\rm tr}_R \,e^{\,u\cdot {\cal C}} =
\frac{\sum_w (-1)^{|w|} e^{\,u\cdot
w(\lambda_R+\rho_W)}}{\sum_w (-1)^{|w|} e^{\,u\cdot w(\rho_W)}}\ ,
\eea
where
$\lambda_R$ is the highest weight, $\rho_W\equiv \sum_{\alpha\in\Delta_+}\alpha/2$
is the Weyl vector, and the sums on the right hand side are over the
Weyl group.
It follows that $ \chi_R(e^{u})= \chi_R(e^{w(u)}) $
for any
Weyl reflection $w$ and thus
\bea
\sum_{\rho\in \mathfrak R} \rho\cdot w(u)= \sum_{\rho\in \mathfrak R} \rho\cdot u\ ,
\eea
Since $u$ is arbitrary and since this holds for any Weyl
reflection $w$, it follows
\bea
\sum_{\rho\in \mathfrak R} \rho^a=0
\eea
for each irreducible representation  $\mathfrak R$ of $\cG$.}

On the other hand, a very different kind of solutions also exist.
Suppose we consider a regime where $x$ scales to zero, along
with $q\rightarrow 0$. For example let us look for solutions
near $x_a^{n_a}\sim q$. At such points $q\rightarrow 0$ might
leave a factor $(1-q/x^{\rho_i}y_i)$ as well. Note that we can
still make use of the infinite product formulae  only if
$q/x^{\rho_i}\rightarrow 0$ as well, so one must first shift
the argument $\rho_i\cdot u$ by an integral multiple of $\tau$,
using the identity
\bea
\theta(w;\tau)=(-1)^m t^{m}q^{m^2/2}\theta(w+m\tau;w) \ ,
\eea
bringing
\bea\label{shift}
&&\Psi(w,\tau) = (-1)^{m} \Psi({}\{ w/\tau {}\}\tau;\tau) \cr\cr
&&= (-1)^{m} e^{-\pi i\tau  {}\{ w/\tau{}\}^2+\pi i\tau {}\{ w/\tau{}\} }
q^{-1/12}\frac{1}{\prod_{k\ge 0}(1-{}\{ t{}\} q^k)(1-q^{k+1}/{}\{ t {}\})} \ ,
\eea
where ${}\{ t{}\}\equiv e^{2\pi i \tau{}\{ w/\tau {}\}} $
with ${}\{ w/\tau {}\}$ obtained by an additive shift
of $w$ by $-m\tau$ for an integer $m$;
the real part of $ {}\{ w/\tau{}\}$ lies between 0 and 1.
The aim is to make the infinite product well-defined
in the limit of $q\rightarrow 0$.
At generic values of $w$, each of the infinite
products reduces to 1, which shows that there is no solution
to 4d BAE, $\Phi_a=1$.

The new type of solutions will have to cluster around special
places, called $H$-saddles, where the real part of $(\rho_i\cdot u+\nu_i)/\tau$
for $\rho_i$'s are integral and not necessarily zero. In the large
$\tau$ limit, with $\nu_i$ kept finite, such $H$-saddles, say $u_H$,
are located at
\bea
\lambda_i\cdot u_H\simeq m_{\lambda_i}\tau
\eea
for some subset $\lambda_i$'s of charge vectors $\rho_i$'s and
accompanying integers $m$'s.
We can then invoke the language of Wilson-line symmetry breaking
and consider $u_H$ as the point where $\lambda_i$'s is
nearly massless while the others are heavily gapped. The holonomy
in question is along the direction 1, which was the fiber circle.

In the neighbourhood of $u_H$, we write for the nearly massless ones
\bea
\lambda_i\cdot u +\nu_i  = m_{\lambda_i}\tau + \lambda_i\cdot \sigma+\nu_i\ ,
\eea
with $\sigma $ and $\nu_i$ understood to remain finite
while $\tau\rightarrow i\infty$. In such a neighborhood of $u_H$,
then, we introduce new shifted Cartan variables $\sigma_a\equiv (u-u_H)_a$ such that
\bea
x^{\lambda_i^a}_a= q^{m} z^{\lambda_i^a}_a, \qquad z_a\equiv e^{2\pi i \sigma_a}\ .
\eea
The shift of $u_a$ to $\sigma_a$ is integral in $\tau$, so can at most
change the sign of $\Psi$ and $\Phi_a$'s, and for these light fields,
we can proceed exactly the same way as the naive limit as in (\ref{vanilla}).
For heavy fields, from $\hat\rho$'s, on the other hand, the infinite
product for the relevant $\theta$ reduces to 1, and leaves behind
a prefactor only.  Then, near $u_H$,  contributions to $\Phi_a$ from
light and heavy modes accumulate to
\bea
\Phi_a (u;\tau)& \rightarrow & \Phi_a^{3d;H}\equiv (-1)^{M}\Lambda_H \prod_i\prod_{\lambda_i}
\left[z^{-\lambda_i/2}y_i^{-1/2}-z^{\lambda_i/2}y_i^{1/2}\right]^{-\lambda^{a}_i}
\eea
for some integer $M$, with
\bea
\label{eq:Lambda_H}
\Lambda_H^a &\equiv &\prod_i\prod_{\hat\rho_i}
e^{-\pi i\tau \hat\rho_i^a \left({}\{
(\hat\rho_i\cdot u_H+\hat \rho_i\cdot \sigma+\nu_i)/\tau{}\}^2
- {}\{  (\hat\rho_i\cdot u_H+\hat \rho_i\cdot \sigma +\nu_i)/\tau{}\} \right)}\ ,
\eea
where we used $\Lambda_G=1$.

As such, the solutions to the 4d BAE  are neatly decomposed into
union of the 3d BAE solutions at these distinct $H$-saddles,
\bea
\{\;u_*\; :\; 1=\Phi_a(u_*)\;\} _{\tau\rightarrow i\infty} \quad =\quad \cup_{u_H} \{\;u_H+\sigma_*\; :\; 1= \Phi_a^{3d;H}(\sigma_*)\;\} \ ,
\eea
where $\lambda_i\cdot u_H/\tau$ are real and integral for some
subset $\lambda_i$'s of $\rho_i$'s. For actual admissible BAE
vacua, we must exclude those solutions that are fixed under some
Weyl reflection but it is clear that this does not interfere with
this classification into clusters around $H$-saddles.
There are two special subclasses of $H$-saddles that are noteworthy.
One is when $\{\lambda\}=\{\rho\}$, which occurs when the gauge holonomy
 is trivial. This of course corresponds to the naive
dimensional reduction, where the 4d BAE reduces to $1=\Phi_a^{3d}$
above. That is, we  included the
first type of solutions on equal footing as well, by assigning
them to $u_H=0$. The other classes are when the unbroken matter charges
$\lambda_i$'s do not span the entire weight vector space. This
means that the reduced 3d theory includes pure gauge sectors
with no chiral fields coupled. The latter deserves a more
in-depth discussion which is postponed to the end of this section,
as this requires a more explicit evaluation of $\Lambda_H$'s.

\subsection{$H$-Saddles in the Small $\tau$ Limit }

We start with the identity,
\bea
\theta(w,\tau)= ie^{-\pi i w^2/\tau}\theta(w/\tau,-1/\tau)\ ,
\eea
which implies that
\bea
\Psi(w,\tau)= \frac{i}{\theta(w/\tau,-1/\tau)}
=\frac{i}{\theta(\tilde w,\tilde\tau)}\ ,
\qquad \tilde w\equiv \frac{w}{\tau}\ ,
\quad \tilde \tau\equiv -\frac{1}{\tau}\ .
\eea
As such, the BAE in the small $\tau$ limit can be cast as
\bea
1=\Phi_a (u,\nu;\tau) \equiv
 \prod_i\prod_{\rho_i} \tilde\Psi(\rho_i\cdot \tilde u+\tilde \nu_i;\tilde \tau)^{\rho^a_i}\ ,
\eea
with
\bea
\tilde \Psi(w,\tau)= ie^{-\pi i \tilde w^2/\tilde\tau}\frac{1}{\theta(\tilde w,\tilde\tau)}\ ,
\eea
since the quadratic exponents in the prefactor will
cancel out for $\Phi_a$ thanks to gauge and axial anomaly cancelation.
This way, the small $\tau$ analysis will
follow the large $\tau$ analysis almost verbatim.

Again, there are two types of solutions. The first class comes with
finite $\tilde u$'s, or equivalently,
\bea
u \;\sim\; \tau, \qquad {\rm as \quad }\tau \;\rightarrow \;i0^+\ ,
\eea
which corresponds to a trivial holonomy along the time circle.
In this obvious saddle, all chiral fields acquire finite mass $\sim \nu_i$ and
thus are in equal footing. The 4d BAE reduces to, as $\tilde q\rightarrow 0$,
\bea\label{vanilla2}
 1=\tilde\Phi_a^{3d}\equiv i^{\#}  \prod_i\prod_{\rho_i}
\left[\tilde x^{-\rho_i/2}\tilde y_i^{-1/2}
-\tilde x^{\rho_i/2}\tilde y_i^{1/2}\right]^{-\rho_i^a} \ ,
\eea
which is essentially the same equation as (\ref{vanilla}) of the large
radius limit, once we replace $x$, $y$, and $q$ by $\tilde x$, $\tilde y$, and $\tilde q$.

Similarly to the large $\tau$ limit, more solutions appear as we
allow $\tilde u$ to scale with the
large $\tilde\tau=-1/\tau$, such that,
\bea
\lambda_i\cdot \tilde u = m_{\lambda_i}\tilde\tau +\lambda_i\cdot \tilde \sigma \ ,
\eea
again exactly as before, for some proper subset $\lambda_i$'s
of $\rho_i$'s, or equivalently
\bea
\lambda_i\cdot u \simeq  -m_{\lambda_i}+\tau \lambda_i \cdot \tilde \sigma\ .
\eea
As in the previous large $\tau $ case, the subset $\lambda_i$'s
represents light degrees of freedom among the matter fields.
Shifting the tilded variables similarly, vacua around such
an $H$-saddle solve
\bea
1=\tilde\Phi_a^{3d;H}\equiv (-1)^{M'/2} \tilde \Lambda_H \prod_i\prod_{\lambda_i}
\left[\tilde z^{-\lambda_i/2}\tilde y_i^{-1/2}-\tilde z^{\lambda_i/2}\tilde y_i^{1/2}\right]^{-\lambda^{a}_i}
\eea
for some integer $M'$, with
\bea
\tilde \Lambda_H^a &\equiv& \prod_i\prod_{\hat \rho_i}
e^{-\pi i\tilde \tau \hat\rho_i^a \left({}\{  (\hat\rho_i\cdot \tilde u_H
+\hat \rho_i\cdot \tilde \sigma+\tilde \nu_i)/\tilde \tau{}\}^2 - {}\{
(\hat\rho_i\cdot \tilde u_H+\hat \rho_i\cdot \tilde \sigma +\tilde \nu_i)/\tilde \tau{}\} \right)} \ .
\eea
As before, the latter ignores overall the phase factor.

The solutions to the 4d BAE can be again grouped into
\bea
\{\; u_*\; :\; 1= \Phi_a( u_*)\;\}_{\tau\rightarrow i0^+} \quad =\quad \cup_{\tilde u_H} \{\;\tilde u_H+\tilde \sigma_*\; :\; 1= \tilde \Phi_a^{3d;H}(\tilde \sigma_*)\;\} \ ,
\eea
where, again, $\lambda_i\cdot \tilde u_H/\tilde\tau $ are real and
integral for some subset $\lambda_i$'s of $\rho_i$'s. As before,
the necessary exclusion of those solutions  fixed under some Weyl
reflection should be performed, which does not interfere with
this $H$-saddle classification of solutions.  The sum
includes the special case of $\{\lambda_i\}=\{\rho_i\}$, where
the 3d BAE is nothing but (\ref{vanilla2}). Note that
the locations of $H$-saddles are along direction 2 in this small $\tau$ limit
while  they were along direction 1 in the large $\tau$ limit.
It is reasonably clear that the solutions to the BAE can be matched,
between the large radius limit and the small radius limit, one on one and
saddle by saddle. As before, the naive $H$-saddle at $u_H=0$ as well
as those that involve pure gauge sector should be included.
See Section \ref{pure}.

\subsection{3d UV Couplings }

In both limits, the naive
reduction on either circle gives 3d BAE, $1=\Phi^{3d}$ or $1=\tilde \Phi^{3d}$,
whereby one recovers vacua with negligible holonomies. This by itself does
not count all 4d BAE vacua, however. In order to account for all vacua,
one must consider the possibility of
turning on some nontrivial holonomies, leading to $1=\Phi^{3d;H}$ near $u_H$
in the large radius limit, or $1=\tilde \Phi^{3d;H}$ near $\tilde u_H$ in the
small radius limit. The holonomy in question is along the circle 1 in the large radius
limit, hence along the fiber circle, and along the circle 2 in the small radius
limit, hence along the time circle, respectively. Either way, the effective
3d theory at a given $H$-saddle comes with reduced field content: only those
associated with weights $\lambda_i$'s remain ``light", while the rest acquire
large masses of order $\beta_{2}/\beta_{1}$ or $\beta_1/\beta_2$, respectively.

When we consider a particular $H$-saddle and 3d effective theory sitting
there, the effect of the heavy modes can manifest via induced
couplings.\footnote{Although we refer to induced 3d couplings at
$H$-saddles here, the computation is straightforwardly extended to
arbitrary holonomy values. At $H$-saddles, one typically considers
split $\rho$'s to $\lambda$'s for light modes and $\hat \rho$'s
for heavy modes, and the UV contribution comes from the latter.
At generic holonomy, however, $\{\lambda\}=\emptyset$ and the
contribution comes from all charged chiral multiplets. }
In 3d theories, the auxiliary $D$-term shows up as
\bea
(\zeta +\sigma \cdot \kappa +\mu\cdot \kappa^F )\cdot D
\eea
with the FI constant $\zeta$, the gauge Chern-Simons level $\kappa$, and the
gauge flavor-mixed Chern-Simons level $\kappa^F$ . $\sigma$ the real scalar in the Cartan
part of the 3d vector multiplet while $\mu$ is the real masses associated
with the flavor symmetries.

This means that one-loop of chiral multiplet of charge $Q$ and
an effective mass $M(\sigma)=Q\cdot \sigma+ M_0$ will induce a shift \cite{Intriligator:2013lca},
\bea
\Delta(\zeta +\sigma \cdot \kappa  +\mu\cdot \kappa^F)=\frac12\, Q\, |Q\cdot \sigma + M_0| \ .
\eea
In the $\tau=\tau_1+i\tau_2\rightarrow i\infty$ limit, the heavy modes have masses $M_0$
of order $|\tau_2|\gg |\sigma|$, and the leading terms
will induce $\zeta\sim |\tau|$ as
\bea
\zeta+ \sigma\cdot \kappa +\mu\cdot \kappa^F
&=&\frac12 \sum_i \sum_{\hat\rho_i} \hat \rho_i\sum_{n\in {\mathbb Z}}\;
|\,{\rm Im}\left(n\tau + \hat\rho_i\cdot u_H  +\hat\rho_i\cdot\sigma +\nu_i\right)|\cr\cr
&+&\frac12 \sum_i \sum_{\lambda_i} \lambda_i^a\sum_{n\in{\mathbb Z} }\;
|\,{\rm Im}\left(n\tau+\lambda_i\cdot\sigma+\nu_i\right)| \ .
\eea
Regularizing the sum, we find, as $\tau\rightarrow i\infty$,
\bea
&&\zeta^a+ (\sigma\cdot \kappa +\mu\cdot \kappa^F)^a\cr\cr
&&\simeq \frac{\tau_2}{2} \sum_{i,\rho_i}
\rho_i^a\left( \bar\epsilon_{\rho_i}+{\rm Im}(\rho_i\cdot \sigma+\nu_i)/\tau_2)-
(\bar\epsilon_{\rho_i}+{\rm Im}(\rho_i\cdot \sigma+\nu_i)/\tau_2)^2\right)\cr\cr
&&\simeq \frac{\tau_2}{2} \sum_{i,\rho_i}
\left( \bar\epsilon_{\rho_i}-\bar\epsilon_{\rho_i}^2\right)\rho_i^a
+\frac{1}{2} \sum_{i,\rho_i} (1-2\bar\epsilon_{\rho_i})\rho_i^a \,{\rm Im}( \rho_i\cdot\sigma+F_i\cdot\nu) \ ,
\eea
where  $\bar\epsilon_{\rho_i}\equiv {}\{ \rho_i \cdot u_H/\tau{}\}$ so that
$\bar\epsilon_{\lambda_i}=0$.
Repeating the exercise for the small $\tau$ limit, we find
\bea
&&\tilde \zeta^a+ (\tilde\sigma\cdot \tilde \kappa +\tilde\mu\cdot \tilde \kappa^F)^a\cr\cr
&&\simeq \frac{\tilde\tau_2}{2} \sum_{i,\rho_i}
\left( \bar\epsilon_{\rho_i}-(\bar\epsilon_{\rho_i})^2\right)\rho_i^a
+\frac{1}{2} \sum_{i,\rho_i}
(1-2\bar\epsilon_{\rho_i})\rho_i^a \,{\rm Im}( \rho_i\cdot\tilde \sigma+F_i\cdot\tilde\nu ) \ ,
\eea
while $\tilde \nu=\nu/\tau$ as with others and
$\bar\epsilon_{\rho_i}\equiv{}\{ \rho_i \cdot \tilde u_H/\tilde \tau {}\}$.
We use the common symbol $\bar\epsilon_\rho$
on the large and the small radius limits since these
two sets of numbers  are really identical.

In either expressions, we can infer the UV contributions to these
couplings in 3d sense, by expanding in $1/\tau_2$ ($1/\tilde\tau_2$)
and dropping $\lambda_i$ contributions
in the second sums for the Chern-Simons level, e.g.,
\bea
&&\zeta^a_{\rm UV}+ (\sigma\cdot \kappa_{\rm UV} +\mu\cdot \kappa^F_{\rm UV})^a\cr\cr
&&\simeq \frac{\tau_2}{2} \sum_{i,\hat\rho_i}
\left( \bar\epsilon_{\hat\rho_i}-\bar\epsilon_{\hat\rho_i}^2\right)\hat\rho_i^a
+\frac{1}{2} \sum_{i,\hat\rho_i} (1-2\bar\epsilon_{\hat\rho_i})\hat\rho_i^a \,{\rm Im}( \hat\rho_i\cdot\sigma+F_i\cdot\mu)
\eea
with $\mu=\nu$. For the small $\tau$ limit, we take  $\mu=\tilde\nu$
and replace $\sigma$ in favor of $\tilde \sigma$.
Although the KK mode sums associated with the charge $\lambda$ could have
contributed to $\zeta_{\rm UV}$, they cancel against the same contributions
from $\hat\rho$'s, thanks to the observation we made earlier, $\sum\rho=0$,
for each irreducible representation for any semi-simple group.
With this, it is clear that these are precisely the couplings
responsible for the prefactor $\Lambda_H$ and $\tilde\Lambda_H$,
\bea
\Lambda_H^a \simeq e^{-2\pi ( \zeta_{\rm UV}+\sigma\cdot\kappa_{\rm UV})^a}\ ,\qquad
\tilde \Lambda_H^a \simeq e^{-2\pi ( \tilde \zeta_{\rm UV}+\tilde \sigma\cdot\tilde \kappa_{\rm UV})^a}\  .
\eea
provided that the left hand sides are appropriately expanded in $1/\tau_2$ ($1/\tilde\tau_2$)
and truncated to the leading order.

The FI constant $\zeta_{\rm UV}$ must be present only for Abelian
part of the subgroup $\cH$ left unbroken by the holonomy $u_H$, which
we wish to confirm as a consistency check. Clearly this would hold
if $\alpha\cdot \zeta_{\rm UV}=0$ for any root $\alpha$ that belongs to
the unbroken groups $\cH$. Since the symmetry breaking to $\cH$ is
due to the holonomy, this means that the irreducible representation
$\mathfrak R_i$ of ${\cal G}$ will decompose into various spin $s$
representations under $SU(2)_\alpha \subset \cH \subset \cG$, and that
\bea
\{\bar\epsilon_{\rho}\;|\;\rho\in \mathfrak R_i\}\quad\rightarrow \quad
\{\bar\epsilon_{l}\;|\; \mathfrak R_i=\oplus_l[s_l]\} \ .
\eea
At the holonomy such that $\bar\epsilon_\alpha=0$, therefore, we have
\bea
\alpha\cdot\sum_{\rho\in \mathfrak R_i} \rho (\bar\epsilon_{\rho} -\bar\epsilon_\rho^2)\;\;
=\;\;\sum_{s_l}(\bar\epsilon_{l} -\bar\epsilon_l^2)\sum_{\mu\in[s_l]}\alpha \cdot (\mu+\cdots)\ ,
\eea
where $\bar\epsilon_l$ denotes the common value of those $\bar\epsilon_\rho$'s
that fall into the $l$-th  irreducible representation, say with spin
$s_l$, of $SU(2)_\alpha$. $\mu$'s are the weights of spin $[s_l]$
representation, embedded into those of $\mathfrak R_i$, and the
ellipsis denotes the part invariant under $SU(2)_\alpha$.
We thus find
\bea
\alpha\cdot\zeta_{\rm UV}\;\;
=\;\;\alpha\cdot\left(\sum_{l}(\bar\epsilon_{l} -\bar\epsilon_l^2) \sum_{\mu\in [s_l]}\mu \right)\;\;=\;\;0 \ ,
\eea
as expected, where in the last step we again used
$\sum_{\mu\in \mathfrak R}\mu=0$ for any irreducible
representation $\mathfrak R$ of a (semi-)simple Lie group.
Generalization to the entire set of $\alpha$ with $\bar\epsilon_\alpha=0$
is immediate.

Also, the Chern-Simons coefficients should be appropriately
quantized. Indeed, we find the gauge Chern-Simons levels
in the UV,
\bea\label{CS}
\kappa_{\rm UV}^{ab}&=& \frac{1}{2} \sum_{i,\hat\rho_i}
\hat\rho_i^a \hat \rho^b_i (1-2\bar\epsilon_{\hat\rho_i}) \cr\cr
&=& \frac{1}{2} \sum_{i,\rho_i} \rho_i^a  \rho^b_i
(1-2\bar\epsilon_{\rho_i})-\frac{1}{2} \sum_{i,\lambda_i} \lambda_i^a  \lambda^b_i
\cr\cr
&=&\frac{1}{2} \sum_{i,\rho_i} \rho_i^a  \rho^b_i (1+2\lfloor \rho_i\cdot u_H/\tau\rfloor)
-\frac{1}{2} \sum_{i,\lambda_i} \lambda_i^a  \lambda^b_i
\eea
holds since $\bar\epsilon_\lambda=0$ and since the gauge anomaly cancelation
demands $\sum_{i,\rho_i}\rho^a_i\rho^b_i\rho^b_i=0$. Recall that
$\lfloor \cdots\rfloor$ means the real integral part, as in (\ref{floor}).
All quantities in the sums are manifestly integral,  so the induced UV Chern-Simons
coefficients are integral up to the overall factor 1/2. Exactly the same
applies to $\tilde \kappa_{\rm UV}$.

The factor 1/2, which some may find troublesome, is not a problem
at all. For many theories, such as the SQCD type where the
fundamental chirals has to appear in pairs, we expect
that this is countermanded by the 4d spectrum.
More to the point, the half-quantized Chern-Simons coefficient is usually
an indication that we must be more careful about the effective action coming from
integration out massive fermions. The usual statement that this leads to Chern-Simons
action is known to miss the global structure of the effective action; whenever the
Chern-Simons coefficient generated is half-integral, and thus apparently variant under large
gauge transformation, the effective action is actually an eta-invariant with full
gauge invariance \cite{AlvarezGaume:1984nf,Witten:2015aba}.

\subsection{Locating $H$-Saddles with Pure Gauge Sectors}\label{pure}
\label{sec:locating H}

So far we have pretended that $H$-saddle would come with
charge vectors $\{\lambda \vert \bar\epsilon_\lambda=0\}$,
enough of them to span the entire charge vector space.
However, this needs not be the case in general, as one
can have 3d theories with unbroken supersymmetry when
the Chern-Simons is nontrivial. Also a further issue arises
when the unbroken gauge group $\cH$ contains a $U(1)$ factor
with a UV FI constant generated. In the latter cases,
some topological vacua may appear shifted far away from
$\sigma,\tilde\sigma\sim O(1)$, potentially muddying the
classification of the $H$-saddles. Here we wish to address
issues related to such $H$-saddles.

Consider the case where $\cH$ contains no Abelian sector. Let us write
\bea
\cH=\cdots \oplus \cK \oplus \cdots \ ,
\eea
where $\cK$ is a semi-simple Lie group with no light 3d chiral
multiplet coupled. Where would such an unbroken group be found? Recall that the
location of the $H$-saddle was determined, so far, by the condition
\bea
\bar\epsilon_\lambda=0
\eea
for some subset of matter charges, $\lambda$'s. In this
current case, no such $\lambda$ knows about $\cK$. Instead the
location of the $H$-saddle is determined by the spontaneous
symmetry breaking as
\bea
\bar\epsilon_\alpha=0,\qquad \alpha \in\mathfrak{K}\ ,
\eea
where $\mathfrak{K}$ is the Lie algebra of $\cK$. When
$\cH$ contains no Abelian subgroup, this combination of
$\{\alpha\}\cup\{\lambda\}$ should span the entire weight
space of $\mathfrak{G}$, the Lie algebra of $\cG$, and
again determine the acceptable positions $u_H$ discretely.

In the absence of light matter fields coupled to $\cK$,
and since no UV FI constant would exist for such non-Abelian
group, the 3d supersymmetric vacua in question are all
``topological" types \cite{Intriligator:2013lca}.
For $\cK=Sp(r)$, for example, one can take a simple basis for the Cartan $U(1)^r$
such that $\sigma = \sum_1^r \sigma_s\cC_s$ with chiral fields in
the defining representations have unit charges with respect to one
and only one $\cC_s$. As such, the reduced BAE will take the simple form, after
some rescaling
\bea
1=(z_s)^{2\kappa_{\rm UV}^{Sp(r)}}\ , \qquad s=1,\cdots ,r \ .
\eea
We remove solutions with $z_s=\pm 1$ for some $s$ or those with
$z_s=z_t$ for some $s\neq t$, and identify those related by Weyl
transformations, $W=S_r\times ({\mathbb Z}_2)^r$. Then the vacua are labeled
by unordered distinct $r$ phases, $e^{\pi i n/\kappa_{\rm UV}^{Sp(r)}}$,
with $1\le n < \kappa_{\rm UV}^{Sp(r)}$
\bea
\left(\begin{array}{c} |\kappa_{\rm UV}^{Sp(r)}|-1 \\ r \end{array}\right) \ .
\eea
Similarly, for $\cK=SU(r+1)$, a simple choice is
$\sigma = \sum_1^r \sigma_s\cC_s-(\sum_s \sigma_s)\cC_{r+1}$
in the redundant basis, whereby the reduced BAE equation becomes
\bea
1=(z_1/z_{r+1})^{\kappa_{\rm UV}^{SU(r+1)}}=\cdots=(z_r/z_{r+1})^{\kappa_{\rm UV}^{SU(r+1)}}
\eea
with $z_{r+1}\equiv [\prod_1^{r} z_s]^{-1}$ understood. Each
equation yields $|\kappa_{\rm UV}^{SU(r+1)}|-1$ acceptable solutions, $z_s/z_{r+1}\neq 1$,
upon which we further impose $z_s/z_{r+1} \neq z_t/z_{r+1}$
for all pairs $s<r$ as well. In the end, the number of acceptable
Weyl-inequivalent solutions is, again
\bea\label{3dYMCS}
\left(\begin{array}{c} |\kappa_{\rm UV}^{SU(r+1)}|-1 \\ r \end{array}\right)\ .
\eea
These dovetail precisely with the 3d index computation by
Witten \cite{Witten:1999ds} once we extend the latter to $\cN=2$; the only new ingredient for
$\cN=2$ is to take $\kappa'=\kappa_{\rm UV}-h$, where $h$ is the dual Coxeter number,
instead of $\kappa'=\kappa_{\rm UV}-h/2$, for the bosonic theory in the end,
since the adjoint fermion content is doubled between the two.
In both cases, therefore, a necessary condition for the existence
of an $H$-saddle involving a pure $Sp(r)$ sector or a pure $SU(r+1)$
sector is $\kappa_{\rm UV}\ge r+1$.

Now let us allow $\cK$ to include an Abelian factor. Unbroken
$U(1)$'s can come with large 3d FI constants as we saw in the
previous section, which will interfere with reduction of 4d BAE to
3d BAE.  If one started with 4d chiral multiplets in real or
pseudo-real representations, such FI constants would cancel out
exactly, but of course this need not be the case.
If one finds large FI constants, say at  some $u_H$, that scale as
Im$\tau$ (or Im$\tilde\tau$), the reduced 3d BAE of such $U(1)\subset\cK$
cannot be solved for $\sigma$ (or $\tilde\sigma$) kept finite.
What this really means is that $H$-saddles must be looked for,
with the condition of $\zeta_{\rm UV}=0$ imposed simultaneously, i.e.,
\bea\label{A}
\zeta^s\;\propto\;\sum_{i,\hat \rho_i}\hat\rho_i^s ( \bar\epsilon_{\hat \rho_i}-\bar\epsilon_{\hat\rho_i}^2)
=\sum_{i,\rho_i}\rho_i^s ( \bar\epsilon_{\rho_i}-\bar\epsilon_{\rho_i}^2)
\quad\rightarrow\quad 0 \ .
\eea
Recall that for this case, neither a matter charge $\lambda$ nor
a root $\alpha$ can be invoked to fix the location of $u^a$. Instead,
we have this $\zeta^s=0$ condition, again fixing the holonomy
$u_H$ to discrete possibilities.

Once this necessary condition is met and the candidate location
for the $H$-saddle is found, we must decide whether
such an $H$-saddle will actually contribute, i.e. whether
the reduced 3d BAE admits nontrivial vacua nearby. Because we are
dealing pure $\cN=2$ gauge theory in three dimensions, the
latter is possible only if $\kappa_{\rm UV}\neq 0$.
 At a saddle with decoupled $U(1)_a$ unbroken group
the actual supersymmetric vacua are determined by the 3d BAE,
\bea\label{Abelian}
C_{s}=(z_s)^{\kappa_{\rm UV}^s}
\eea
for some finite constant $C_{U(1)}$, so  the number of them is
\bea
|\kappa_{\rm UV}^s| \ .
\eea
Therefore, we have found that an $H$-saddle involving a decoupled
$U(1)_s$ gauge sector is possible provided that $\zeta_{\rm UV}^s=0$
and  $\kappa_{\rm UV}^s\neq 0$.

The question of $H$-saddles with $\cK= U(1)\subset \cH$ but now
with light 3d charged matter field, a general case of (2) above, is a
little more involved. Suppose we located a candidate $H$-saddle
using a condition of type $\bar\epsilon_\lambda=0$ for some $U(1)$-coupled
charge vector $\lambda$. A schematic form of the rank 1
3d BAE at such an $H$-saddle is
\bea\label{Abelian with matter}
1\;\simeq\; q^{\xi} z^{\kappa - \sum Q^2/2+\sum (Q')^2/2}
\frac{\prod_Q (y z^{Q}- 1)^{Q}}{\prod_{Q'} ( z^{Q'}-y)^{Q'}}\ ,
\eea
or
\bea\label{far}
\prod_{Q'}( z^{Q'}-y)^{Q'}\;\simeq\;  q^{\xi} z^{\kappa -\sum Q^2/2+\sum (Q')^2/2}\prod_{Q} (y z^{Q}- 1)^{Q}\ ,
\eea
where $-Q$ and $Q'$ denote, collectively, the light charges of negative and
positive signs respectively, and $\xi\equiv \zeta_{\rm UV}/{\rm Im}\tau$ and
$\kappa=\kappa_{\rm UV}$ or $\xi\equiv \tilde\zeta_{\rm UV}/{\rm Im}\hat\tau$
and $\kappa=\tilde \kappa_{\rm UV}$, in the large or in the small $\tau$ limits,
respectively. Here, we will consider a large $\tau$ limit, or $q\rightarrow 0$,
without loss of generality.

Suppose that $\xi>0$. For $\xi<0$, we get the same result after
flipping  $Q \leftrightarrow Q'$
and $\kappa \leftrightarrow -\kappa$. If $\xi=0$, there is no issue,
to begin with, as all solutions would be $O(1)$ and do not scale with $q$.
Setting $q=0$ for the moment, we find $Q'$ finite solutions $z\sim y^{1/Q'}$
to (\ref{far}), each of which are $Q'$ times degenerate. As we turn back
on small $q$, these would split but remain finite. To enumerate the other
worrisome solutions that scale with $q$ or $1/q$, it is useful define
$k\equiv \kappa-\sum Q^2/2+\sum (Q')^2/2$, $l\equiv \kappa+\sum Q^2/2-\sum(Q')^2/2$.
We then find,

\begin{itemize}

\item $k\ge 0$, $l > 0$,

$l$ large solutions $z\sim q^{-\xi/l}$;

the total number of solutions are $\sum (Q')^2+l=\kappa+\sum Q^2/2+\sum(Q')^2/2$;

\item $k\ge 0$, $l \le 0$,

no new solutions;

the total number is $\sum (Q')^2$;

\item $k<0$, $l \le 0$,

$-k$ small solutions $z\sim q^{\xi/|k|}$;

the total number is $\sum (Q')^2-k=-\kappa+\sum Q^2/2+\sum (Q')^2/2$;

\item $k<0$, $l> 0$,

$l$ large solutions $z\sim q^{-\xi/l}$ and $-k$ small solutions $z\sim q^{\zeta/|k|}$;

the total number is $\sum (Q')^2+l-k=\sum Q^2$ .

\end{itemize}

Among these vacua, the large and the small ones $z\sim q^\#$ should be
taken only as a qualitative indication that somewhere far away there
exist supersymmetric and topological vacua of free $U(1)$ Chern-Simons
theory. The truncation to (\ref{Abelian with matter}) is justified only
at finite values of $\sigma$, and thus precise locations of these additional vacua
should be worked out by going back to the 4d BAE. Why is the reduction to 3d
theory, which has worked flawlessly so far, compromised? Simply because
the dimensional reduction ends up with 3d FI constant which still remembers
the large value of $\tau$ and thus the fact that the purported 3d theory
came from 4d theory with the extremely elongated $T^2$. The number of
vacua found for such $U(1)$ factor in the preceding analysis should
still hold, but the precise locations of those at $z\sim q^\#$ are
not to be trusted. Rather, one must really view this situation as a sum
of distinct $H$-saddles consisting of two types. One is $U(1)$ theory
with charge matters, but with its topological vacua due to very large
FI constant excised. The others are free $U(1)$ theory elsewhere in
the $u$-space, with no light matter field coupled and supersymmetric
vacua, due to Chern-Simons level $k$ or $l$, as in (\ref{Abelian}).

To summarize, locating $H$-saddles involves three sets of data,
\bea
\bar\epsilon_\lambda=0,\qquad \bar\epsilon_\alpha=0,\qquad \zeta_{\rm UV}=0\ ,
\eea
and one proceeds by collecting at least rank-many conditions to
fix discrete locations in the Cartan torus spanned by $u$'s.
In particular, when we end up a $U(1)$ factor coupled with charged
3d matter and large UV FI constant
$\zeta\sim {\rm Im}\tau, \;{\rm Im}\tilde\tau$, we must take care
to discard the far-away topological  vacua $\sigma \sim \zeta$ from
such an $H$-saddle and instead look for a nearby saddle with free
$U(1)$ factor at vanishing FI constant and non-vanishing Chern-Simons level.
With finite chemical potentials and a matter content symmetric under
the charge conjugation, this latter complication never appears.
On the other hand, such $U(1)$ cases will be more typical in the
large radius limit for the so-called ``physical" version, regardless
of matter content, because of large $U(1)_R$ chemical potential.
Next, we now move to this last type of $H$-saddles.

\subsection{$H$-Saddles with Large Chemical Potentials}

As we hinted already, the large radius limit of
the ``physical" $S^1\times S^3$ partition function deviates a little
from the main story of this note. Apart from why this has to be
so from the viewpoint of how these objects are constructed,
we can also trace the difference at a mathematical level to
the large $U(1)_R$ chemical potential $(r-1)\tau$. The latter
shifts the argument of various operators by a large amount
in the large $\tau$ limit, common for each chiral multiplets
in a single irreducible gauge representation.

With non-integral $r$'s, in particular, necessary at the superconformal point,
one immediate consequence is that, even if the 4d theory came
with charge-conjugation symmetric gauge representation, the
light 3d field content, if any, would not be generically so;
the positively charged matter and the negatively charged
matter would become light at different holonomies. At candidate
$H$-saddles,  one
will typically encounter unbroken $U(1)$ gauge theories with
large uncanceled $\zeta_{\rm UV}$, which shifts the location
of the saddle to far away, and makes the search for genuine
$H$-saddle qualitatively different from the other cases. For this
reason, we will denote these rather distinct $H$-saddle values
of $u$'s by introducing the notation $\hat u_H$.

One obvious place to look for a saddle, independent of
details, is $\hat u_H=0$. Here the 4d gauge group $\cG$ will descend
to 3d intact, while chiral multiplets with typical values of $r_i$
will become all massive. As such no FI constant would be generated,
as the 3d gauge group $\cH=\cG$ would have no $U(1)$ factor.
For a qualitative understanding,
we will confine our attention to theories with a single classical Lie
group $\cG$ as the gauge group and further assume that $0<r_i<1$ for
all matter multiplets. Is there an $H$-saddle located at the
naive choice $\hat u_H=0$?

The pure $\cG$ Yang-Mills-Chern-Simons (YMCS) theory there would have no
supersymmetric vacua unless one finds sufficiently large
UV Chern-Simons level, which can be easily computed as,
\bea
\kappa_{\rm UV}^{ab}\;=\;\delta^{ab}\;\gamma_\cG T^{(2)}_{\rm def}\;
\kappa_{\rm UV}^{\cG}\ ,
\eea
where $T^{(2)}_{\rm def}$ is that of the defining representation
and $\gamma_{SO}=1/2$ and $\gamma_{SU}=\gamma_{Sp}=1$.\footnote{
For actual vacuum counting via BAE, however, we restrict ourselves
at most to $SU$ and $Sp$ cases: See the top of Section 2.}
Then, we find
\bea\label{CSorigin}
\kappa_{\rm UV}^{\cG}
\;=\;\frac{1}{2 \gamma_\cG T^{(2)}_{\rm def}}\;\sum_i T^{(2)}_i(1-2 r_i )\ .
\eea
On the other hand, the Adler-Bell-Jackiw (ABJ) anomaly cancelation requires
\bea
T^{(2)}_{\rm adj}\;=\;\sum_i T^{(2)}_i(1- r_i)\, ,
\eea
so that
\bea\label{CSSCI0}
\kappa_{\rm UV}^{\cG}
\;=\;\frac{T^{(2)}_{\rm adj} }{ \gamma_\cG T^{(2)}_{\rm def}}
\times \sum_i \frac{T^{(2)}_i}{T^{(2)}_{\rm adj}} (1/2- r_i )
\;=\;\frac{T^{(2)}_{\rm adj} }{\gamma_\cG T^{(2)}_{\rm def}}\cdot\left(
1-\frac32 \cdot \frac{\sum_i T^{(2)}_i}{3T^{(2)}_{\rm adj}}\right) \, .
\eea
Note that the second term inside the parentheses cannot exceed $3/2$,
once we demand the asymptotic freedom.
If the theory contains a single type of chiral multiplets, the
asymptotic freedom combined with $0<r$ implies
\bea
|\kappa_{\rm UV}^{\cG}|
\; < \;\frac{T^{(2)}_{\rm adj} }{ 2\gamma_\cG T^{(2)}_{\rm def}} =h_\cG \ ,
\eea
with the dual Coxeter number $h_\cG$. Recall that the counting
of 3d vacua for pure YMCS theories is dictated by the difference
between $|\kappa|$ and $h$. This leads us to suspect that, for
all asymptotically free theories that flow to CFT, the naive
$\hat u_H=0$ saddle is absent.

For $SU(N_c)$ theory with $N_f$ fundamental and $N_f$ anti-fundamental
chirals, e.g., (\ref{CSorigin}) gives, since
$r=1-N_c/N_f$ by the ABJ anomaly cancelation,
\bea
\kappa_{\rm UV} = 2N_c-N_f
\eea
for 4d conformal field theories, $3N_c/2\le N_f <3N_c$.
With
\bea
|\kappa_{\rm UV}| < h_{SU(N_c)}=N_c\ ,
\eea
(\ref{3dYMCS}) tells us that the naive
saddle at the origin, $\hat u_H=0$, has no supersymmetric vacua and
thus is not an $H$-saddle. For  SQCD theories, the saddle at
origin is actually absent.
Similar considerations for $Sp(r)$ theory with $2N_f$
fundamental flavors show that, again there is no $H$-saddle
at $\hat u_H=0$ for asymptotically free theories $N_f\le 3r+2$;
for $N_f=3r+2$, one finds $\kappa=-r$.

With no $H$-saddle at the origin, next places are those holonomies
with vanishing UV FI constants,
\bea\label{FISCI}
\sum \rho^a(\hat \epsilon_\rho -\hat \epsilon_\rho^2) \;\;=\;\;0 \ .
\eea
Suppose that the matter content is symmetric under charge conjugation,
such that charge vectors always come in pairs $(\rho,-\rho)$. Then,
places where this happens generically are
\bea
\rho \cdot \hat u_H \;\;\in\;\; \tau{\mathbb Z}/2\ ,
\eea
which allows $\epsilon_{\hat\rho}=\epsilon_{-\hat\rho}$ and thus
pairwise cancelations in the sum (\ref{FISCI}). Assuming $r_i\neq 1/2$,
the theory reduces to pure Yang-Mills type and the Chern-Simons
level is
\bea
\kappa_{\rm UV}^{ab} \;\;=\;\;\delta^{ab}\;\gamma_\cG T^{(2)}_{\rm def}\;
\kappa_{\rm UV}^{\cG} +\sum_{i,\rho_i} \rho^a_i\rho^b_i\;\lfloor \rho_i\cdot \hat u_H/\tau+r_i\rfloor\ ,
\eea
where $\kappa_{\rm UV}^{\cG}$ is the Chern-Simons level at $\hat u_H=0$ as in
(\ref{CSSCI0}). Coming back to $SU(2)$ theory with $2N_f$ fundamental
flavors, we find that reduced theory is a pure $SU(2)$ YMCS with
\bea
\kappa_{\rm UV} = -2, 4,\qquad {\rm for}\;\; N_f=3,5\ ,
\eea
implying 1 and 3 BAE vacua, respectively, which
are consistent with the Witten index of the original
4d theories.

$N_f=4$ with $r=1/2$ at SCFT also admit $\hat u_H=\tau/2$
as an $H$-saddle; the reduced theory is an $SU(2)$
theory with $2N_f=8$ fundamental chirals and vanishing
UV Chern-Simons level. The number of vacua for this
3d theory is usually expected to be three. However, the actual theory at
this $H$-saddle, being a reduction from 4d where the
baryonic $U(1)$ is anomalous, and, because this theory
cannot have UV FI constant, one of these potential vacua
is pushed to the Coulombic infinity. The number of
vacua at the $\hat u_H=\tau/2$ saddle is actually 2
which is again consistent with the 4d Witten index.
These suggest that for SQCD theories, $\hat u_H=\tau/2$
is the only $H$-saddle in the Casimir limit.

\section{4d Theory as a Disjoint Sum of 3d Theories}

These discussions lead us to a clear definition of $H$-saddle
for general supersymmetric gauge theories on a compact spacetime
with a small circle or a small circle bundle. In the small radius
limit, a $d$ dimensional partition function $\Omega_d$ will reduce
to a sum of $(d-1)$ dimensional partition functions, ${\cal Z}_{d-1}^H$,
\bea\label{reduction}
\Omega_d \;\;\rightarrow \;\; \sum_{u_H} c_H {\cal Z}_{d-1}^H
\eea
labeled by some discrete choices of the holonomy including the
trivial one. The prefactors $c_H$ capture contributions from
the Kaluza-Klein towers as well as massive multiplets.
For the partition functions we have been studying,
\bea
\Omega_4^{g;p_1,p_2} = \sum_{u_*} {\cal H}^{g-1}
{\cal F}_1^{p_1}  {\cal F}_2^{p_2}
\eea
with the two circles $(p_1,p_2)$-fibred over genus $g$
surface, we find that this decomposes, both in the large
and in the small $\tau$ limits,
\bea
\Omega_4^{g;p_1,p_2} \;\;\rightarrow\;\; \sum_{u_H}
\left(\sum_{\sigma_*} {\cal H}^{g-1}
{\cal F}_1^{p_1}  {\cal F}_2^{p_2}\right)\ ,
\eea
where the 4d BAE vacua are reorganized into sets of 3d BAE vacua
for mutually disjoint 3d theories at various $H$-saddles. As we
already emphasized, we find such decomposition even in the large
$\tau$ limit, because, in effect, this is equivalent to a small
radius limit of  circle 1.

For $g\neq 1$, we can reorganize the sum over 3d vacua $\sigma_*$
at each $u_H$, in terms of the 3d BAE partition function, ${\cal Z}^H_3$
and a multiplicative factor $c_H$. The latter will generically have
have an exponential behavior, interpreted as the Cardy exponent
in the small $\tau$ limit and as the Casimir energy in the large $\tau$ limit.
Previous estimates of such leading exponents have effectively
considered only the naive saddle at $u_H=0$ \cite{Assel:2015nca,Bobev:2015kza}.
The results from such computations, interpreted as being
related to 4d anomaly polynomials, must be therefore questioned.

\subsection{4d Witten Index is a Sum of 3d Witten Indices}

Before we plunge into quantitative studies, it is worthwhile
to consider the simplest case of $p_{1,2}=0$ and $g=1$. For $T^4$, the
BAE computes the numerical Witten index, with the summand at each
$u_*$ equal to 1. This means that we have an intuitive relation
\bea
{\cal I}_{4}^G\;\; = \;\;
 \sum_{u_H}\;{\cal I}_{3}^H\ ,
\eea
whereby the 4d Witten is reconstructed from those of several 3d
theories sitting at distinct holonomies. Regardless of the details
of computations to follow for different $g$'s and $p$'s, this by
itself tells us that the small radius limit of 4d theory cannot
be regarded as a single 3d theory.
If we are considering supersymmetric theories in
compact spacetime, therefore, the 4d theory in the small
radius limit should be considered as a disjoint sum
of 3d theories.

Since the same set of $u_H$'s
enters such decompositions for all ${\cal M}_4^{g;p_1,p_2}$'s,
this also means that an $H$-saddle will occur if and only
if the reduced 3d theory there has nontrivial Witten index.
The latter condition can be considered as the most important single
property of $H$-saddles. The class of theories we are considering
in this note are maximally mass-deformed by flavor holonomies so
that the partition functions and Witten indices are all integral.
As such, an $H$-saddle would occur if and only if the Witten
index of the reduced 3d theory at the candidate holonomy is
non-vanishing.

On the other hand, the notion of $H$-saddle
clearly goes beyond the particular class of theories or
background geometries we are considering in this note. More
generally twisted partition functions are often not enumerative,
resulting in non-integral twisted partition functions. As we
recalled in the Introduction, twisted partition function
would generally compute the analog of the ``bulk index".
In such cases, the defining property of the $H$-saddle
should be extended to allow non-vanishing supersymmetric
partition function of the reduced theory at the candidate
holonomy.

If we were considering the 4d theory on $S^1\times \mathbb R^3$,
the holonomies would label superselection
sectors; the dimensional reduction process is
ambiguous until we specify the holonomy or compute the vacuum
expectation value of the holonomy. The above relation tells us
that there are discrete choices of $u_H$ whereby the dimensional
reduction produces distinct 3d theories whose 3d supersymmetry is
not spontaneously broken, and that the 4d Witten index is reproduced
only after we sum over the Witten indices of these 3d theories at
distinct $u_H$'s.

Such a behavior of 4d theory on a circle,
producing multiple 3d theories in the small radius limit, has been noted
previously by Seiberg and collaborators while studying how 4d
dualities reduce to 3d dualities \cite{Aharony:2013dha}. As the above relation
shows, a dual pair of 4d theories would produce, each,
several 3d theories which must be collectively dual to
each other. Whether or not this implies individual
3d dualities, say, in our language at $H$-saddle pairs,
is in principle another matter. For 4d theory as a
starting point, however, the interpretation of $u_H$ as
the superselection sector label does suggest that the duality
will hold for 3d theories pairwise, or in our
terminology, $H$-saddle by $H$-saddle.

The robust nature of the Witten index under
small deformations is often invoked to simplify index
computations. One such would be the insensitivity to
the size of the circles in $T^4$, but this, if used
improperly, seems to imply that Witten indices agree
between theories in different dimensions if one is a
dimensional reduction of the other. However, we already
know, via  many examples, that this is not quite correct.
For instance, Witten pointed out how the index of 4d
$\cN=1$ pure Yang-Mills is sensitive to disconnected
sectors of mutually commuting holonomies along $T^3$ \cite{Witten:2000nv};
such sectors would be dropped if one or more
radii of $T^3$ is taken to zero literally.

Our relation is yet another reminder that such topological
invariance argument should not be taken too far.
The problem with the zero radius limit of a spacetime
circle is that the compact space of
holonomies becomes noncompact in a zero radius limit, and
cannot be considered a small deformation.
The above formula, which is far more general than the
particular class of theories here and the partition functions thereof,
gives a neat way to relate Witten indices
of gauge theories in adjacent dimensions.

We close with two simple examples. The first is the
canonical SQCD, namely $SU(N)$ theories  with
$N_f$ fundamental and $N_f$ anti-fundamental chirals. For these,
it is clear that the only $H$-saddle is the one
at $u_H=0$, hence we have
\bea
{\cal I}_{4}^G\;\; = \;\;{\cal I}_{3}^H\biggr\vert_{u_H/\tau=0}\ ,
\eea
where the reduced $H$ theory at $u_H/\tau=0$ has the
same gauge group and the same chiral multiplet content
as its 4d cousin $G$. Indeed,
Closset et. al. \cite{Closset:2017bse} found,
\bea
{\cal I}_{4}^G\;\; = \;\;{\cal I}_{3}^H\biggr\vert_{u_H/\tau=0}\;\;=\;\;
\left(\begin{array}{c} N_f-2 \\ N-1 \end{array}\right)\ .
\eea
Since the matter representation
is real collectively, neither the
FI constant nor the Chern-Simon level arise at UV.

The other, less trivial, example is an $SU(2)$ theory
with two fundamental chirals and two adjoint chirals.
For this, a nontrivial $H$-saddle at
$u_H/\tau=1/2$ is present as well as the
naive one at $u_H=0$. While we are formulating
things via the large radius limit, the small
radius limit is found, verbatim, by replacing the
variables to the tilded ones with the identical
result. As such we have
\bea
{\cal I}_{4}^G\;\; = \;\;\sum_{u_H/\tau=0, 1/2}{\cal I}_{3}^H
\eea
where the 3d theory at $u_H/\tau=1/2$
has the two adjoint chirals only. Again no UV 3d coupling
is generated at either saddle, and 3d BAE vacua can be
counted straightforwardly. We find
\bea
{\cal I}_{4}^G\;\; = \;\;
{\cal I}_{3}^H\biggr\vert_{u_H/\tau=0}
\;+\;{\cal I}_{3}^H\biggr\vert_{u_H/\tau=1/2} \;\;=\;\;8\;+\;6\;\;=\;\;14\ .
\eea
The main feature of the latter example, relative to the first,
is a chiral multiplet with gauge representation beyond the fundamental one.
In fact, the existence of chiral multiplet in a gauge
representation larger than the defining one, for
classical ones at least, is one obvious criterion
for nontrivial $H$-saddle to exist.

Note that, of these, the first example is not compatible with
general A-twist background, since the anomaly-free
$U(1)_R$ charge is not integral.
For $\Sigma_g=T^2$, however, the A-twist is null, so we do not
need to restrict $U(1)_R$ charge to be integral. And as long
as can find non-anomalous $U(1)_R$, its chemical potential
can be turned on. For this reason, this recursive computation
of the Witten index can be used for more examples of theories
than generic geometries of this class would allow.
Of course, this is up to the major caveat that theories
being considered are all equipped with real masses in
the 3d sense, as is a common downside of the BAE formulation.
One must take care, in general, not to confuse the Witten
index computed this way with those of the vanilla 4d
$\cN=1$ theories on $R^4$.

We close this subsection with a caveat. In relating 4d theories to
one or more 3d theories, obtained by dimensional reduction at such
saddles, we are always speaking of the small radius limit. This means
that constraints from the 4d anomaly, for example, should be considered
valid even in the said 3d limit. One example is the $SU(N_c)$ SQCD,
whose strict 3d form allows an extra $U(1)$ flavor symmetry which
would be anomalous in 4d. Our 3d theories at $H$-saddles are 
the ones without such a global symmetry; this affects the allowed
superpotential and hence the 3d Witten index as well.

\subsection{Asymptotics}

Now let us turn to other, more involved partition functions.
In literature, some 4d partition functions have been discussed with a particular interest on their asymptotic behavior \cite{Ardehali:2015bla,DiPietro:2016ond,DiPietro:2014bca,Bobev:2015kza,Martelli:2015kuk,Assel:2015nca}. Especially, Ardehali first observed the influence of the holonomy on the Cardy limit of the superconformal index \cite{Ardehali:2015bla}, i.e., the partition function on $S^3 \times S^1$, which is later extended to more general manifolds by Di Pietro and Honda \cite{DiPietro:2016ond}. The latter discussed the Cardy limit of the $\mathcal M_3 \times S^1$ partition function, with explicit examples $\mathcal M_3 = L(n,1), \, \Sigma_g \times S^1$.

On the other hand, for the Casimir limit, the role of the holonomy is rarely discussed as far as we are aware. In this section, we provide a unified way of examining both the Cardy limit and the Casimir limit of the partition function, which manifests itself in the $H$-saddle approach.

\subsubsection*{Asymptotics of ${\cal H}$}

The handle-gluing operator is, with $r_i$ being the
$U(1)_R$ charge of  the $i$-th chiral multiplet,
\bea
{\cal H}\equiv \eta(\tau)^{-2\,\mathrm{rank}(\cG)} \prod_\alpha \Psi(\alpha\cdot u; \tau)\prod_i\prod_{\rho_i} \Psi(\rho_i\cdot u+\nu_i;\tau)^{r_i-1}\times {\rm det}\left[\frac{\partial_a \log \Phi_b}{2\pi i}\right]\ ,
\eea
of which the last piece can, at most, contribute logarithmic
corrections in the exponent. The large and the small $\tau$ limit of
$\Psi$'s were already explored. These may be combined to, for the
gauge multiplet contributions, at each $H$-saddle,
\bea
\eta(\tau)^{-2\,\mathrm{rank}(\cG)}\prod_\alpha \Psi(\alpha\cdot u; \tau)\biggr\vert_{\tau\rightarrow i\infty}
\sim\quad q^{-{\rm dim}(\cG)/12} e^{\pi i\tau \sum_\alpha \epsilon_{\alpha}(1-\epsilon_{\alpha})}\ ,
\eea
where $\epsilon_{\alpha}={}\{ \alpha\cdot (u_H+\sigma)/\tau{}\}$
are real numbers between 0 and 1,
\bea
\eta(\tau)^{-2\,\mathrm{rank}(\cG)}\prod_\alpha \Psi(\alpha\cdot u; \tau)\biggr\vert_{\tau\rightarrow i0^+}
\sim \quad \tilde q^{-{\rm dim}(\cG)/12} e^{\pi i\tilde \tau \sum_\alpha  \tilde \epsilon_{\alpha}(1-\tilde \epsilon_{\alpha})}\ ,
\eea
where $ \tilde \epsilon_{\alpha}={}\{\alpha\cdot (\tilde u_H+\tilde\sigma)/\tilde\tau {}\}$
are real numbers between 0 and 1. The chiral multiplet contributions
can be written similarly as
\bea
\prod_i\prod_{\rho_i} \Psi(\rho_i\cdot u+\nu_i;\tau)^{r_i-1}\biggr\vert_{\tau\rightarrow i\infty}
\sim\quad q^{-\sum_i\sum_{\rho_i}(r_i-1)/12} e^{\pi i\tau \sum_i(r_i-1)\sum_{\hat\rho_i} \epsilon_{\hat\rho_i}(1-\epsilon_{\hat\rho_i})}\ ,
\eea
and
\bea
\prod_i\prod_{\rho_i} \Psi(\rho_i\cdot u+\nu_i;\tau)^{r_i-1}\biggr\vert_{\tau\rightarrow i0^+}
\sim\quad \tilde q^{-\sum_i\sum_{\rho_i}(r_i-1)/12} e^{\pi i\tilde \tau \sum_i(r_i-1)\sum_{\hat\rho_i}\tilde \epsilon_{\hat\rho_i}(1-\tilde \epsilon_{\hat\rho_i})}\ ,
\eea
where $\epsilon_{\rho_i}={}\{ (\rho_i\cdot (u_H+\sigma)+\nu_i)/\tau{}\}$
and $\tilde \epsilon_{\rho_i}={}\{(\rho_i\cdot (\tilde u+\tilde \sigma)+\tilde\nu_i)/\tilde\tau{}\}$ are also
real numbers between 0 and 1.

\subsubsection*{Asymptotics of ${\cal F}_1$ }

The first fibering operator is given by
\begin{align}
\mathcal F_1 = \prod_i \prod_{\rho_i} \Xi_1(\rho_i \cdot u+\nu_i;\tau)
= \prod_i \prod_{\rho_i} e^{2 \pi i \left(\frac{(\rho_i \cdot u+\nu_i)^3}{6 \tau^2}
-\frac{\rho_i \cdot u+\nu_i}{12}\right)} \Gamma_0 (\rho_i \cdot u+\nu_i;\tau)\ ,
\end{align}
where
\begin{align}
\Gamma_0 (u;\tau) = \prod_{n = 0}^\infty \left(\frac{1-x^{-1} q^{n+1}}{1-x q^{n+1}}\right)^{n+1}\ .
\end{align}
To find the large radius limit of $\mathcal F_1$, again we decompose
$\rho_i \cdot u+\nu_i$ into $(\epsilon_{\rho_i}+m_{\rho_i}) \tau$
where $\epsilon_{\rho_i}$ belongs in the range $0 \leq \epsilon_{\rho_i} < 1$ and $m_{\rho_i}$
is the integer part. Using
\begin{align}\label{Cascade}
\Xi_1 (u+m \tau;\tau) = e^{-\frac{\pi i}{2} (m^2+m)} \Psi (u;\tau)^{-m} \Xi_1 (u;\tau)
\end{align}
for an integer $m$, one can find the large radius limit of $\mathcal F_1$ as follows:
\begin{align}\label{S3Cardy}
 \mathcal F_1\biggr\vert_{\tau \rightarrow i \infty} \sim \prod_i \prod_{\rho_i}
 e^{\pi i \tau \left(\frac{\epsilon_{\rho_i}^3}{3}
 +\epsilon_{\rho_i}^2 m_{\rho_i}-\epsilon_{\rho_i} m_{\rho_i}-\frac{\epsilon_{\rho_i}}{6}+\frac{m_{\rho_i}}{6}\right)}\ ,
\end{align}
with $\rho_i \cdot u+\nu_i = (\epsilon_{\rho_i}+m_{\rho_i}) \tau$.

The identity (\ref{Cascade}) also resolves an apparent puzzle with this asymptotic
formula. Note that under a large gauge transformation $u_a$ can be shifted to $u_a+\tau$.
This will induce shift of both $\epsilon_\rho$'s and $m_\rho$'s, under which
the exponent of (\ref{S3Cardy}) does not look particularly invariant.
Let us first look at how $\mathcal F_1$ transforms. Since $\rho\cdot u$ will shift by
$\rho_a\tau$, the transformation is
\bea
\mathcal F_1  \;\;\rightarrow\;\; \left[\prod_{i}\prod_{\rho_i} (-1)^{\rho^a_i}
\Psi^{-\rho^a_i}\right]\times \mathcal F_1 = \Phi_a^{-1}\times \mathcal F_1 \ ,
\eea
where we used $\sum_\rho\rho^a=0$ for each irreducible representations,
as was shown in Section 3.2. What this formula tells us is that
although the flux operator ${\cal F}_1$ is not invariant as a function of $u$
under such large gauge transformations, its values at supersymmetric vacua, where
$1=\Phi_a$, are invariant. Therefore, although the leading exponent in
(\ref{S3Cardy}) may look odd, its values at $H$-saddles are really
invariant under $u_a\rightarrow u_a+\tau$. The same kind of invariance
will work for ${\cal F}_2$ under $u_a\rightarrow u_a+1$, for the small
$\tau$ limit, as the two are related by $S$-transformation.

On the other hand, the small radius limit of $\mathcal F_1$ can be obtained using the S-transformation. First note that $\Xi_1$ satisfies an identity \cite{Closset:2017bse}
\begin{align}
\label{eq:S}
\Xi_1 (u;\tau) = e^{\frac{\pi i}{\tau^2} \frac{u^3}{3}} \Xi_2 \left(\frac{u}{\tau};-\frac{1}{\tau}\right)\ ,
\end{align}
where $\Xi_2$ is defined by
\begin{align}
\Xi_2 (u;\tau) = e^{2 \pi i \left(\frac{u^3}{6 \tau}-\frac{u^2}{4}+\frac{u \tau}{12}+\frac{1}{24}\right)} \prod_{k = 0}^\infty \frac{f(u+k \tau)}{f(-u+(k+1)\tau)}\ ,
\end{align}
and
\begin{align}
f(u) = \exp \left[\frac{1}{2 \pi i} \mathrm{Li}_2 (e^{2 \pi i u})+u \log \left(1-e^{2 \pi i u}\right)\right]\ .
\end{align}
$\Xi_2$ satisfies
\begin{align}
\label{eq:T}
\Xi_2 (u+m \tau;\tau) = e^\frac{\pi i m}{6} \Xi_2 (u;\tau)\ ,
\end{align}
and, as a result, $\Xi_1$ can be written as
\begin{align}
\Xi_1 (\rho_i \cdot u+\nu_i;\tau) = e^{-\pi i \tilde \tau^2 \frac{(\tilde \epsilon_{\rho_i}+\tilde m_{\rho_i})^3}{3}+\frac{\pi i (\tilde m_{\rho_i})}{6}} \Xi_2 \left(\tilde \epsilon_{\rho_i} \tilde \tau;\tilde \tau\right)\ ,
\end{align}
where $\rho_i \cdot\tilde u+\tilde \nu_i$ is decomposed into $(\tilde m_{\rho_i}+\tilde \epsilon_{\rho_i})\tilde \tau$
with $0 \leq \tilde \epsilon_{\rho_i} < 1$ and an integer $\tilde m_{\rho_i}$. The cubic phase term will vanish after summed over all the multiplets due to the anomaly-free condition.
$f(u)$ comes from the 1-loop determinant of a chiral multiplet on $S^3$ and converges to 1 for large $u$,
\begin{align}
\label{eq:limit}
 f(u)\biggr\vert_{u \rightarrow i \infty} = 1\ .
\end{align}
Thus, $\mathcal F_1$ has the following asymptotic behavior for $\tau \rightarrow i 0^+$:
\begin{align}
\label{eq:F}
 \mathcal F_1\biggr\vert_{\tau \rightarrow i 0^+} \sim \prod_i \prod_{\rho_i} e^{\pi i\tilde\tau^2 \left(\frac{\tilde \epsilon_{\rho_i}^3}{3}-\frac{\tilde \epsilon_{\rho_i}^2}{2}+\frac{\tilde \epsilon_{\rho_i}}{6}\right)\ .
}
\end{align}
\subsubsection*{Asymptotics of ${\cal F}_2$}

The second fibering operator $\mathcal F_2$ is given by
\begin{align}
\mathcal F_2 &= \prod_i \prod_{\rho_i} \Xi_2 (\rho_i \cdot u+\nu_i) \\
&= \prod_i \prod_{\rho_i} e^{2 \pi i \left(\frac{(\rho_i \cdot u+\nu_i)^3}{6 \tau}-\frac{(\rho_i \cdot u+\nu_i)^2}{4}+\frac{(\rho_i \cdot u+\nu_i) \tau}{12}+\frac{1}{24}\right)} \, \prod_{k = 0}^\infty \frac{f(\rho_i \cdot u+\nu_i+k \tau)}{f(-\rho_i \cdot u-\nu_i+(k+1) \tau)}\ .
\end{align}
In the large radius limit,
we find the following limit of $\mathcal F_2$:
\begin{align}
 \mathcal F_2\biggr\vert_{\tau \rightarrow i \infty} = \prod_i \prod_{\rho_i} e^{\pi i \tau^2 \left(\frac{\epsilon_{\rho_i}^3}{3}-\frac{\epsilon_{\rho_i}^2}{2}+\frac{\epsilon_{\rho_i}}{6}\right)}\ ,
\end{align}
with $\rho_i \cdot u+\nu_i = (\epsilon_{\rho_i}+m_{\rho_i}) \tau$.  Similarly one can also find the small radius limit of $\mathcal F_2$ using the S-transformation \eqref{eq:S}:
\begin{align}
 \mathcal F_2\biggr\vert_{\tau \rightarrow i 0^+} = \prod_i \prod_{\rho_i} e^{-\pi i\tilde \tau \left(\frac{\tilde \epsilon_{\rho_i}^3}{3}+\tilde \epsilon_{\rho_i}^2  m_{\rho_i}- \tilde \epsilon_{\rho_i} \tilde m_{\rho_i}-\frac{ \tilde \epsilon_{\rho_i}}{6}+\frac{m_{\rho_i}}{6}\right)}\ ,
\end{align}
where $\rho_i \cdot\tilde u+\tilde \nu_i=(\tilde m_{\rho_i}+\tilde \epsilon_{\rho_i})\tilde \tau$
with $\tilde \epsilon_{\rho_i} \in [0,1)$. Note that
the small and the large $\tau$ limits of ${\cal F}_2$ mirror,
under $\tau\rightarrow -1/\tau$,
the large and the small $\tau$ limits of ${\cal F}_1$, faithfully and respectively.

\subsubsection*{Asymptotics of $\mathcal F^\text{phys}$}

Unlike the A-twist gauge, the physical handle-gluing
operator only contains Jacobian factor $H^\text{phys}$,
which does not contribute to the leading term of the
partition function. The leading contribution then only
comes from $\mathcal F^\text{phys}$. From \eqref{eq:F}
one can see that each component of $\mathcal F^\text{phys}$
has the following asymptotic behavior.

In the small $\tau$ limit, with
\begin{align}
\tilde \epsilon'_{\rho_i} &= {}\{ (\rho_i \cdot (u_H+\tilde \sigma)+\tilde \nu_i)/\tilde \tau+l_R (r_i-1) /\tilde \tau {}\}\ , \\
\tilde \epsilon'_{\alpha} &= {}\{ \alpha \cdot (u_H+\tilde \sigma)/\tilde \tau+l_R/\tilde \tau {}\}\ ,
\end{align}
we find similarly
\begin{align}
\prod_i \prod_{\rho_i \in \mathfrak R_i} \Xi_1(\rho_i \cdot u+\nu_i+l_R \, \tau (r_i-1);\tau)\biggr\vert_{\tau \rightarrow i 0^+} \sim \prod_i \prod_{\rho_i} e^{\pi i\tilde\tau^2 \left(\frac{(\tilde \epsilon'_{\rho_i})^3}{3}-\frac{(\tilde \epsilon'_{\rho_i})^2}{2}
+\frac{\tilde \epsilon'_{\rho_i}}{6}\right)\ ,
},
\end{align}
and
\begin{align}
(-1)^{\frac{l_R (l_R+1)}{2} \mathrm{rank}(\cG)} \eta(\tau)^{2 l_R \mathrm{rank}(\cG)} \prod_{\alpha} \Xi_1(\alpha \cdot u+l_R \, \tau;\tau)\biggr\vert_{\tau \rightarrow i 0^+} \sim \prod_{\mathfrak G} e^{\pi i\tilde\tau^2 \left(\frac{(\tilde \epsilon'_{\alpha})^3}{3}-\frac{(\tilde \epsilon'_{\alpha})^2}{2}+\frac{\tilde \epsilon'_{\alpha}}{6}\right)\ ,
},
\end{align}
where the last product is taken over all the gauge generators, again.
It is important to note here that the set of $H$-saddles
and the subsequent values of $\bar\epsilon$'s to be used in
the subsequent expansion of the exponents are
no different from the preceding discussion of the small
$\tau$ limit of A-twisted
cases. This happens because the shift due to $\nu_R$ is
negligible as $\tau\rightarrow i0^+$, as far as the
values of $u_H$ are concerned. Clearly this is not
the case for the other limit $\tau\rightarrow i\infty$.

As we noted already in Section 3, the large radius limit
$\tau\rightarrow i\infty$ for ``physical" cases follows
a different pattern due to the large shift $\nu_R = l_R \, \tau = \frac{1-g}{p} \, \tau$. With
\begin{align}
\epsilon'_{\rho_i} &= {}\{ (\rho_i \cdot (\hat u_H+\sigma)+\nu_i)/\tau+l_R (r_i-1) {}\}\ , \\
\epsilon'_{\alpha} &= {}\{ \alpha \cdot (\hat u_H+\sigma)/\tau+l_R {}\}\ ,
\end{align}
where $m'_{\rho_i}, \, m'_\alpha$ are the remaining integer parts,
the exponential behavior goes as
\bea
&&  \prod_i \prod_{\rho_i \in \mathfrak R_i}
\Xi_1(\rho_i \cdot u+\nu_i+l_R \, \tau (r_i-1);\tau)\biggr\vert_{\tau \rightarrow i \infty} \cr\cr
&&\quad \sim \prod_i \prod_{\rho_i} e^{\pi i \tau \left(\frac{(\epsilon'_{\rho_i})^3}{3}+(\epsilon'_{\rho_i})^2 m'_{\rho_i}-\epsilon'_{\rho_i} m'_{\rho_i}-\frac{\epsilon'_{\rho_i}}{6}+\frac{m'_{\rho_i}}{6}\right)}\ ,
\eea
and
\bea
&&  (-1)^{\frac{l_R (l_R+1)}{2} \mathrm{rank}(\cG)} \eta(\tau)^{2 l_R \mathrm{rank}(\cG)} \prod_{\alpha} \Xi_1(\alpha \cdot u+l_R \, \tau;\tau)\biggr\vert_{\tau \rightarrow i \infty} \cr\cr
&&\quad \sim \prod_{\mathfrak G} e^{\pi i \tau \left(\frac{(\epsilon'_{\alpha})^3}{3}+(\epsilon'_{\alpha})^2 m'_{\alpha}-\epsilon'_{\alpha} m'_{\alpha}-\frac{\epsilon'_{\alpha}}{6}+\frac{m'_{\alpha}}{6}\right)}\ ,
\eea
where the last product is taken over all generators of the
gauge group with $\epsilon=0$ understood for the Cartan generators.

\subsubsection*{$H$-Saddles from the Small Radius Limit of a Fibred Circle }

We close with a minor consistency check on the notion of $H$-saddle
by considering the collapsing circle with nontrivial winding number
$p$ over the base. The Cardy limit $\beta_1\rightarrow 0$ with $p_1\neq 0$
would be the canonical example, while the Casimir limit
$\beta_1\rightarrow \infty$ with $p_2\neq 0$ shares the same
issue since as far as our partition functions go this
is equivalent to the other Cardy limit $\beta_2\rightarrow 0$.
The winding number $p$ of the collapsing circle is not
part of 3d spacetime data, so should not enter the 3d
partition functions ${\cal Z}^H_3$'s, since the notion of
$H$-saddle relies on the existence of 3d theories that
makes sense without referring to its 4d origin. It would
be allowed to enter the coefficients $c_H$'s which serve
as the glue between the 3d theories at $H$-saddles and
the original 4d theory.

In view of the lengthy discussions in Section 3, it should be
relatively clear that the part of $({\cal F}_1)^{p_1}$ that could have
contributed to ${\cal Z}^H_3$ in the large $\tau$ limit resides
entirely in $\Gamma_0$ of Eq.~(\ref{Gamma0}). However the latter function reduces
to 1 universally as $q\rightarrow 0$, regardless of the field content.
The winding number $p_1$ therefore
contributes at most to $c_H$'s, in this limit, via the surviving
exponential prefactors in front of $\Gamma_0$'s, and does not
interfere with the 3d theory at the $H$-saddles. Then,
$SL(2,{\mathbb Z})$  automatically implies that the
same happens for the $\beta_2\rightarrow 0$ limit with $p_2\neq 0$,
as ${\cal F}_2$ in the small $\tau$ limit is nothing but
${\cal F}_1$ in the large $\tau$ limit modulo exponential
prefactors, which are again harmless for the issue here.

For physical cases, one should look at how
$\mathcal F^\text{phys}$ behaves in the $\tau\rightarrow i\infty$ limit.
The relevant part of the latter fibering operator is made up
of $\Xi_1$'s, or $\Gamma_0$'s therein, so again,
$\mathcal F^\text{phys}$ reduces to a product of exponential
functions: The winding number $p$ can contribute to $c_H$'s
at most, again as promised.

\subsection{Cardy and Casimir}

In the small and the large $\tau$ limits, we found
exponential behaviors of the partition function which
differ between different $H$-saddles.
The partition function on A-twisted geometries, for
example, always has an $H$-saddle at $u_H=0$, and
the exponential behavior there follows a universal
form,
\bea
{\cal H}^{g-1}\biggr\vert_{u_H=0} \quad\sim\quad \left[e^{2\pi i\tau \cdot (-{\rm tr}_f R)/12} \right]^{g-1}
\quad{\rm or}\quad \left[e^{2\pi i\tilde\tau \cdot (-{\rm tr}_f R)/12}\right]^{g-1}\ ,
\eea
in the respective limits of large $\tau$ or $\tilde\tau=-1/\tau$.
${\rm tr}_f R$ means the trace of $U(1)_R$ charge over all 4d
fermions. The exponents inferred from this universal
part have been identified in the past and given interpretation
of the Casimir energy and the Cardy exponent with respect to
the large and the small radius limit of $\beta_2$ \cite{Closset:2017bse}.
The same expression also appeared for the Cardy limit of SCI's,
which was then related to  conformal anomaly coefficients
\cite{DiPietro:2014bca}.

Existence of $H$-saddles at $u_H\neq 0$ and the different
leading exponents at such places, however, tell us that
the Cardy exponent and the Casimir energy may be rather
different in general. In this last section, we will
explore this issue. For the sake of simplicity we will
confine our attention to pure imaginary $\tau=i\beta_2/\beta_1$
and consider only those 4d theories whose chiral field
content is invariant under the charge conjugation symmetry,
$\rho\rightarrow -\rho$.

\subsubsection*{A-Twist}

We find, at each $H$-saddle at $u_H$,
the leading exponents of ${\cal H}^{g-1}$ in the large $\tau$ limit
is
\bea\label{aH}
(g-1)\times\left[ \,-\frac{1}{12}({\rm tr}_f R)
+\frac12 \sum_\alpha \epsilon_{\alpha}(1-\epsilon_{\alpha})+
\frac12\sum_i(r_i-1)\sum_{\rho_i} \epsilon_{\rho_i}(1-\epsilon_{\rho_i})\,\right]
\eea
multiplied by $2\pi i\tau$, instead of the universal form
\bea\label{naiveCardy}
(g-1)\times\left[\, -\frac{1}{12}({\rm tr}_f R)\;\right]
\eea
at $u_H=0$.
Clearly the $H$-saddle with the dominant contribution and the
exponent thereof may be identified only after comparing
this expression at different $H$-saddles. Furthermore,
the actual exponent is given by this multiplied by $(g-1)$, so
the dominant contributions for $g=0$ and the dominant contributions
for $g>1$ will generically come from different $H$-saddles.
For the small $\tau$ limit, the same formulae work with $\epsilon$'s and $\tau$
replaced by $\tilde\epsilon$'s and $\tilde \tau$.

Note that, once we begin to identify
3d BAE vacua and evaluate the sum, $\epsilon$'s at a given
$H$-saddle would be really
\bea
\epsilon_\rho=\bar\epsilon_\rho+ \frac{\rho_i\cdot \sigma_* +\nu_i}{\tau}\,\quad\hbox{or}\quad
\tilde \epsilon_\rho = \bar\epsilon_\rho+\frac{\rho_i\cdot \tilde \sigma_* +\tilde \nu_i}{\tilde \tau}
\eea
etc, for multiple $\sigma_*$'s found by solving the 3d BAE
at $u_H$. Expanding (\ref{aH}), the leading term
\bea
(g-1)\times\left[ \,-\frac{1}{12}({\rm tr}_f R)
+\frac12 \sum_\alpha \bar\epsilon_{\alpha}(1-\bar\epsilon_{\alpha})+
\frac12\sum_i(r_i-1)\sum_{\rho_i} \bar\epsilon_{\rho_i}(1-\bar\epsilon_{\rho_i})\,\right]
\eea
must be augmented by the sub-leading pieces
\bea\label{subH}
\frac{(g-1)}{\tau}\times\left[ \sum_\alpha \bar\epsilon_\alpha \,\alpha
+\sum_i(r_i-1)\sum_{\rho_i}\bar\epsilon_{\rho_i}\,\rho_i \right]\cdot\sigma_* \ , \cr\cr
-\frac{(g-1)}{\tilde\tau}\times\left[ \sum_\alpha \bar\epsilon_\alpha \,\alpha
+\sum_i(r_i-1)\sum_{\rho_i}\bar\epsilon_{\rho_i}\,\rho_i \right]\cdot \tilde \sigma_* \ ,
\eea
which, combined with the overall factor $2\pi i\tau$ ($2\pi i\tilde \tau$),
supply finite and $\sigma_*$ ($\tilde\sigma_*$) dependent
phases. Thus, cancelations between 3d BAE vacua in favor
of smaller exponents at a given $H$-saddle cannot be
ruled out in general. Although such cancelations do
not appear to be commonplace, we will identify a few examples
of this kind later.

For $p_{1,2}\neq 0$, there is a further exponential contribution
of the form, via ${\cal F}_1^{p_1}{\cal F}_2^{p_2}$ in the
sum. In the large $\tau$ limit, the additional terms,
to be added to (\ref{aH}), are
\bea\label{pF1}
&&p_1\times\left[\sum_i \sum_{\rho_i}
\frac12 \left(\frac{\epsilon_{\rho_i}^3}{3} +\epsilon_{\rho_i}^2 m_{\rho_i}-\epsilon_{\rho_i} m_{\rho_i}-\frac{\epsilon_{\rho_i}}{6}+\frac{m_{\rho_i}}{6}\right)\right]\cr\cr
&+&p_2\times\left[\sum_i \sum_{\rho_i} \frac{\tau}{2}
\left(\frac{\epsilon_{\rho_i}^3}{3} -\frac{\epsilon_{\rho_i}^2}{2}+\frac{\epsilon_{\rho_i}}{6}\right)\right]\ ,
\eea
again modulo the large multiplicative factor $2\pi i\tau$.
For the small $\tau$ limit, we merely need to
exchange the asymptotic forms of $\cF_1$ and of $\cF_2$
and replace $\epsilon_{\rho_i} \rightarrow \tilde \epsilon_{\rho_i}, \, m_{\rho_i} \rightarrow \tilde m_{\rho_i}$ and $\tau\rightarrow \tilde\tau$.

In particular, with the restriction of the matter
content to be symmetric under the charge conjugation,
all terms that involve $\rho\cdot u$ cancel away
leaving behind those involving powers of $\nu_i$'s.
The above then reduces to, e.g. for the large $\tau$ limit,
\bea\label{pF1}
&&p_1\times\left[\sum_i \frac{\nu_i}{\tau} \sum_{\rho_i}
\frac12 \left(\bar\epsilon_{\rho_i}^2 +2\bar\epsilon_{\rho_i} m_{\rho_i}- m_{\rho_i}-\frac{1}{6}\right)\right]\cr\cr
&+&p_2\times\left[\sum_i \nu_i\sum_{\rho_i} \frac{1}{2}
\left(\bar\epsilon_{\rho_i}^2 -\bar\epsilon_{\rho_i}+\frac{1}{6}\right)\right]\ ,
\eea
the latter of which contributes  $\nu$-dependent pieces to
the Casimir energy, while the former $1/\tau$ term contributes
a finite imaginary piece to the exponent.

\subsubsection*{Physical}

For the ``physical" case, the $H$-saddle behavior is different
between the large radius limit and the small radius limit, as
we saw in the previous subsection. The small radius limit
itself is on par with that of A-twisted case, except that
only ${\cal F}^{\rm phys}$ contributes the leading exponential
\begin{align}
\label{matterp}
p\times\sum_i \sum_{\rho_i \in \mathfrak R_i} \left[\frac{\tilde\tau }{2} \left(\frac{\bar \epsilon_{\rho_i}^3}{3}-\frac{\bar \epsilon_{\rho_i}^2}{2}+\frac{\bar \epsilon_{\rho_i}}{6}\right)
+\frac{1}{2} \left[\rho_i \cdot \tilde \sigma+\tilde \nu_i+l_R (r_i-1)\right] \left(\bar \epsilon_{\rho_i}^2-\bar \epsilon_{\rho_i}+\frac{1}{6}\right)\right] \ ,
\end{align}
from matters, which reduces to
\bea\label{matterp'}
\sum_i \sum_{\rho_i \in \mathfrak R_i} \left[p \times \frac{\tilde \nu_i}{2} \left(\bar \epsilon_{\rho_i}^2-\bar \epsilon_{\rho_i}+\frac{1}{6}\right)
+(1-g) \times \frac{r_i-1}{2} \left(\bar \epsilon_{\rho_i}^2-\bar \epsilon_{\rho_i}+\frac{1}{6}\right)\right]\ ,
\eea
on theories with matter content which is symmetric under charge
conjugation. The contribution from the vector is
\bea\label{vectorp}
&& (1-g) \times \sum_{\mathfrak G}\left[\frac{1}{2} \left(\bar \epsilon_{\alpha}^2-\bar \epsilon_{\alpha}+\frac{1}{6}\right)\right]\ ,
\eea
Both appear in the exponent with $2\pi i\tilde\tau$ multiplied.

The expression (\ref{matterp'}) plus (\ref{vectorp}), with $\tilde\nu_i $ set
to zero, has been isolated for the high-temperature
limit of the 4d superconformal index \cite{Ardehali:2015bla}, i.e., $p=1$ and $g=0$,
and govern the asymptotic behavior of the integrand prior to the
holonomy integration, called $V_{\rm eff}$ as in Ref.~\cite{Ardehali:2015bla} modulo
a constant shift.
One subtlety is that the expressions we found via the BAE
are  meant to be evaluated and used at discrete places, $u=u_H$'s,
so agreement with Ref.~\cite{Ardehali:2015bla} requires that the maximum of $V_{\rm eff}$
necessarily occurs at an $H$-saddle. In fact, this is very likely since
$V_{\rm eff}$ is a piece-wise linear function, as a consequence of ABJ
anomaly cancelation, and the derivative changes only at points
where one or more charged fields become massless. Thus the local maximum
and minimum can only occur at places where $q\cdot u \in {\mathbb Z}$
for some charge $q$, and for a full agreement we only need to exclude places,
$u_0$ where a vector multiplet of charge $\alpha$ becomes massless
and no chiral multiplets are.

Since the contributions of the chiral multiplets to $V_\text{eff}$,
after using the anomaly condition, cannot change abruptly there
and since contribution from the $\alpha$-charged vector will make
a sharp turn, it suffices to consider how the derivative of
\begin{align}
V_\alpha = -\epsilon_\alpha^2+\epsilon_\alpha-\frac{1}{6}
\end{align}
behaves at $\epsilon_\alpha=0$, i.e., where $\alpha \cdot u$ becomes
an integer. Let's consider a small neighborhood around $u_0$
parameterized by $-1< t <1$ as $u = u_0+t v$ where $v$ is an
arbitrary direction. Depending on the sign
of $\alpha \cdot v$, the integer part of $\alpha \cdot u = n_\alpha+t \,
\alpha \cdot v$ changes from $n_\alpha-1$ to $n_\alpha$ or $n_\alpha$ to
$n_\alpha-1$ as $t$ crosses $t = 0$. Thus, the vector multiplet
contribution turns sharply at $t = 0$. Computing the derivatives before
and after, it is easy to see that the turn
\begin{align}
 \frac{d V_\alpha}{d t}\biggr\vert_{t = 0^+}- \frac{d V_\alpha}{d t}
\biggr\vert_{t = 0^-} = 2 |\alpha \cdot v|
\end{align}
is positive for arbitrary $v$. The point $u_0$ cannot be a local
maximum,
which means that the maximum of $V_{\rm eff}$ cannot occur at
such a point. It would occur at one of $H$-saddles, therefore,
which gives a full agreement on the Cardy
exponent between the previous approach and the BAE.

One can also deduce the large radius limit of the ``physical'' case. Recall, for each BAE solution, the leading exponent from $\mathcal F^\text{phys}$ is given by
\begin{align}
\label{eq:physCasimir}
& p \times\left[\sum_i \sum_{\rho_i \in \mathfrak R_i}
\frac12 \left(\frac{(\epsilon'_{\rho_i})^3}{3} +(\epsilon'_{\rho_i})^2 m'_{\rho_i}-\epsilon'_{\rho_i} m'_{\rho_i}-\frac{\epsilon'_{\rho_i}}{6}+\frac{m'_{\rho_i}}{6}\right)\right. \nonumber \\
&\qquad \left.+\sum_{\mathfrak G}
\frac12 \left(\frac{(\epsilon'_\alpha)^3}{3} +(\epsilon'_\alpha)^2 m'_\alpha-\epsilon'_\alpha m'_\alpha-\frac{\epsilon'_\alpha}{6}+\frac{m'_\alpha}{6}\right)\right]\ ,
\end{align}
with $2 \pi i \tau$ multiplied. Since the partition function in total is obtained by summing up the contributions with those leading exponentials, the simplest guess would be that the Casimir energy equals the smallest exponent among the values of \eqref{eq:physCasimir} evaluated at the BAE solutions.

\subsubsection*{Cancelations in the Casimir Limit}

However, we also encounter a large class of examples where
the Casimir energies do not equal the smallest exponents
computed above. The primary examples are found in
the superconformal indices, i.e., the partition function
in ``physical"  background
with $p = 1, \, g = 0$. The  Casimir energy of the
resulting SCI's turns out to to be equal to the value of
\eqref{eq:physCasimir}at $\hat u_H = 0$, despite the presence
of nontrivial $H$-saddles. This holds, in many cases
for SCI, even with the naive $\hat u_H=0$ saddle  absent.

This surprising fact can be demonstrated by rewriting
the partition function as a unit circle contour integral,
\begin{align}\label{unitcircle}
\sum_{u_* \in S_\text{BE}} \mathcal F^\text{phys}(u_*,\nu;\tau) H(u_*,\nu;\tau)^{-1} = \frac{1}{|W_G|} \int_{|x| = 1} \frac{dx}{2 \pi i x} \mathcal F^\text{phys}(x,y;q)\ ,
\end{align}
the leading factor of $\mathcal F^\text{phys}$, which
would have generated the holonomy-dependent Casimir
energy,
\begin{align}\label{constantphase}
& \left(\prod_i \prod_{\rho_i \in \mathfrak R_i} e^{\frac{\pi i}{3 \tau^2} [\rho_i \cdot u+\nu_i+\tau (r_i-1)]^3-\frac{\pi i}{6} [\rho_i \cdot u+\nu_i+\tau (r_i-1)]}\right) \left(\prod_\mathfrak G e^{\frac{\pi i}{3 \tau^2} (\alpha \cdot u+\tau)^3-\frac{\pi i}{6} (\alpha \cdot u+\tau)}\right) \nonumber \\
& = e^{\sum_i \dim (\mathfrak R_i) \left[\frac{\pi i}{3 \tau^2} (\nu_i+\tau (r_i-1))^3-\frac{\pi i}{6} (\nu_i+\tau (r_i-1))\right]+\frac{\pi i}{6} \tau \dim (\mathfrak G)}
\end{align}
becomes independent of $u$ due to anomaly conditions.
Thus, it comes out of the integral, and the leading
exponent of $q$ is fixed by the value of \eqref{eq:physCasimir}
at $\hat u_H = 0$ \cite{Bobev:2015kza}. When we come back to
BAE form, this happens via numerous cancelations
between BAE vacua and sometimes even between
$H$-saddles.

This cancelation is possible in part because positions
of $H$-saddles are aligned along the real axis of $2\pi i u$
in this case.  This should be contrasted to the Cardy limit,
where the $H$-saddles are located along the unit circle
$|x|=1$ so that $H$-saddle phenomena manifests even in
this alternate integral formula. When the $H$-saddle
occurs along the unit circle $|x|=1$, the cancelations
due to the anomaly cancelation condition no longer works
because the infinite product formula must be rewritten
in new shifted variables whenever one crosses such $H$-saddle; this was at the
heart of the $H$-saddle computation. The previous observation by Ardehali on
Cardy exponents \cite{Ardehali:2015bla} has effectively captured this $H$-phenomenon on such a unit
circle version of the superconformal indices. In contrast,
such a cancelation does not happen in the Cardy limit.

Something similar happens for the Casimir limit of
the A-twist case when the fibration is nontrivial. To see this, we should keep
the finite part of the leading terms of $\mathcal F$ and $\mathcal H$.
For simplicity, we focus on the rank-1 case with $p_1 = p, \ p_2 = 0$.
For massive matter fields at a given $H$-saddle, $u_H$, the contribution from $\mathcal F$ is
\begin{align}
\prod_i \prod_{\rho_i} e^{\pi i \tau \left(\frac{\bar \epsilon_{\rho_i}^3}{3}+\bar \epsilon_{\rho_i}^2 m_{\rho_i}-\bar \epsilon_{\rho_i} m_{\rho_i}-\frac{\bar \epsilon_{\rho_i}}{6}+\frac{m_{\rho_i}}{6}\right)+\pi i \left(\bar \epsilon_{\rho_i}^2+2 \bar \epsilon_{\rho_i} m_{\rho_i}-m_{\rho_i}-\frac{1}{6}\right) (\rho_i \sigma+\nu_i)-\frac{\pi i}{2} (m_{\rho_i}^2+m_{\rho_i})}\ .
\end{align}
For massless matter fields, i.e., for $\rho_i = \lambda_i$ such that $\epsilon_{\lambda_i} = 0$, we have an additional factor
\begin{align}
\times (1-z^{\lambda_i} y_i)^{m_{\lambda_i}}
\end{align}
with $z = e^{2 \pi i \sigma}$.
Using $m_{\lambda_i} = \lambda_i u_H/\tau$, the contribution from massless fields can be written as follows:
\begin{align}
& \prod_i \prod_{\lambda_i} e^{-\frac{\pi i}{2} (\lambda_i^2 u_H^2/\tau^2+\lambda_i u_H/\tau)} q^{\lambda_i u_H/\tau/12} z^{-\lambda_i/12} y_i^{-1/12} \left[z^{-\lambda_i/2} y_i^{-1/2}-z^{\lambda_i/2} y_i^{1/2}\right]^{\lambda_i u_H/\tau} \nonumber \\
&= \prod_i \prod_{\lambda_i} e^{-\frac{\pi i}{2} (\lambda_i^2 u_H^2/\tau^2+\lambda_i u_H/\tau-2 M u_H/\tau)} q^{\lambda_i u_H/\tau/12} z^{-\lambda_i/12} y_i^{-1/12} \Lambda_H^{u_H/\tau} \left(\Phi^{3d;H}_a\right)^{-u_H/\tau}\ ,
\end{align}
where $M = \sum_i \sum_{\rho_i} m_{\rho_i} \rho_i$ and $\Lambda_H$ is defined in \eqref{eq:Lambda_H}. Note that $u_H/\tau$ is a rational number in $[0,1)$. Thus, at $\sigma = \sigma_*$, the last factor becomes a root of unity,
\begin{align}
 \left(\Phi^{3d;H}_a\right)^{-u_H/\tau}\biggr\vert_{\sigma = \sigma_*} = e^{-2 \pi i k u_H/\tau}\ .
\end{align}
$\mathcal H$ consists of two parts: $e^{2 \pi i \Omega}$ and $H$. For massive matter fields, the leading term of $e^{2 \pi i \Omega}$ is given by
\begin{align}
\prod_i \prod_{\rho_i} e^{\pi i \tau (r_i-1) \left(-\bar \epsilon_{\rho_i}^2+\bar \epsilon_{\rho_i}-\frac{1}{6}\right)+\pi i (r_i-1) \left(-2 \bar \epsilon_{\rho_i}+1\right) (\rho_i \sigma+\nu_i)+\pi i (r_i-1) m_{\rho_i}}\ ,
\end{align}
while for massless matter fields, we have an additional factor
\begin{align}
\times (1-z^{\lambda_i} y_i)^{-(r_i-1)}\ .
\end{align}
The same expansion can be made for vector fields, by replacing $\rho_i \rightarrow \alpha$ and $r_i \rightarrow 2$. Moreover, for $SU(2)$, $H$ is explicitly written as
\begin{align}
& \sum_i \sum_{\rho_i} |\rho_i|^2 \left[\frac{1}{2}-{}\{(\rho_i (\hat u_H+\sigma)+\nu_i)/\tau{}\}\right. \nonumber \\
&\qquad \qquad \qquad \left.+\sum_{k = 0}^\infty \frac{{}\{ x_H^{\rho_i} z^{\rho_i} y_i{}\} q^k}{1-{}\{ x_H^{\rho_i} z^{\rho_i} y_i{}\} q^k}-\sum_{k = 0}^\infty \frac{{}\{ x_H^{-\rho_i} z^{-\rho_i} y_i^{-1}{}\} q^{k+1}}{1-{}\{ x_H^{-\rho_i} z^{-\rho_i} y_i^{-1}{}\} q^{k+1}}\right]\ .
\end{align}
For massive fields, the first line is the leading contribution of order $q^0$ while for massless fields, there is an extra $\mathcal O(q^0)$ contribution
$+\frac{z^{\lambda_i} y_i}{1-z^{\lambda_i} y_i}$. A similar expansion is made for physical gauge as well by replacing $\nu_i \rightarrow \nu_i+\nu_R (r_i-1)$.

\subsubsection*{Explicit Examples with $\cG=SU(2)$: the Casimir Limit}

We now explore some explicit examples for the Casimir limit; recall
that this side is prone to further subtleties beyond $H$-saddles.
Let us discuss the A-twist case first. Considering asymptotically free theories of $SU(2)$, allowed representations are those with isospin $\frac{1}{2} \leq s \leq \frac{3}{2}$. For a model with few number of matters, BAE tends to be trivial due to lack of enough flavor symmetry and cannot be discussed using the A-twist formalism. The Intriligator-Seiberg-Shenker (ISS) model \cite{Intriligator:1994rx} is such an example. It has no anomaly-free flavor symmetry and, as a result, has the fixed anomaly-free $R$-charge $R = 3/5$, which allows A-twist only on a manifold of genus $g \in 5 \mathbb Z$ due to the Dirac quantization condition for $R$-charges. Thus, we relegate the discussion of this model to the physical gauge case, and here consider $SU(2)$ with fundamentals and adjoints.

The $RGG$ anomaly condition restricts $R$-charges of fundamentals and adjoints such that
\begin{align}
\sum_{i = 1}^{N_f} (r_i-1)+4 \sum_{j = 1}^{N_a} (\tilde r_j-1)+4 = 0\ ,
\end{align}
where $r_i$ and $\tilde r_j$ are the $R$-charges of fundamentals and adjoints respectively. For simplicity, we take
\begin{align}
r_i = 1+\frac{4 (N_a-1)}{N_f},\qquad \tilde r_j = 0\ .
\end{align}
With one adjoint, the numbers of flavors allowed by the asymptotically-free condition are $N_f = 2, \, 4, \, 6$. In those cases, however, the exponent \eqref{aH} is independent of $u$, so not very interesting in our discussion. Instead, we discuss the $SU(2)$ model with two fundamentals and two adjoints. Because of the adjoints, the $H$-saddles for this model are located at $u_H/\tau = 0$ and $u_H/\tau = 1/2$.

Take the $H$-saddle at $u_H/\tau = 1/2$. At this $H$-saddle only the vector field and the adjoint matter fields are massless while the fundamental matter fields become massive. Thus, the reduced BAE is given by
\begin{align}
\frac{\prod_{j = 1}^{2} (z^2-w_j)^2}{\prod_{j = 1}^{2} (1-z^2 w_j)^2} = 1\ .
\end{align}
The equation has eight solutions, which are classified into two classes $S_\pm$ satisfying
\begin{align}
\frac{\prod_{j = 1}^{2} (z^2-w_j)}{\prod_{j = 1}^{2} (1-z^2 w_j)}\biggr\vert_{z = z_*} = \pm1, \qquad z_* \in S_\pm \ .
\end{align}
For positive sign, the equation reduces to
\begin{align}
z^4 = 1 \ ,
\end{align}
which has solutions $z = \pm 1, \, \pm i$. Among them, since $z = \pm 1$ are Weyl invariant, only $z = \pm i$ are relevant solutions. For negative sign, on the other hand, the equation can be reorganized into
\begin{align}
\label{eq:S-}
(1+w_1) (1+w_2)= \frac{z^4 (w_1+w_2)-2 z^2 (1+w_1 w_2)+w_1+w_2}{(1-z^2)^2}\ .
\end{align}

The fibering operator and the handle-gluing operator at $u_H/\tau =1/2$
are expanded as follows:
\bea
\mathcal F &= &\frac{e^{\frac{\pi i}{\tau^2} f(\nu_i,\mu_i)}}{w_1^{7/12} w_2^{7/12}} \frac{\prod_{j = 1}^{2} (1-z^2 w_j)}{\prod_{j = 1}^{2} (z^2-w_j)}+\mathcal O(q^\frac{1}{2}), \cr\cr
\mathcal H&=& q^\frac{7}{12}\times
\frac{4 \, e^{\frac{\pi i}{\tau} h(\nu_i,\mu_i)} (1-w_1) (1-w_2) (1-w_1 w_2)}{w_1^{3/2} w_2^{3/2}}\cr\cr
&& \times \frac{z^4 (w_1+w_2)-2 \, z^2 (1+w_1 w_2)+w_1+w_2}{(1-z^2)^2} +\mathcal O(q^\frac{13}{12})\ ,
\eea
where
\begin{gather}
f(\nu_i,\mu_i) = \frac{2}{3} \, \nu_1^3+\frac{2}{3} \, \nu_2^3+\mu_1^3+\mu_2^3, \\
h(\nu_i,\mu_i) = -4 \, \nu_1^2-4 \, \nu_2+3 \, \mu_1^2+3 \, \mu_2^2.
\end{gather}
One immediately notes that the leading term of $\mathcal F$ is proportional to the square root of BAE. Thus,
\begin{align}
&  \mathcal F\biggr\vert_{z = z_*} = \frac{\pm e^{\frac{\pi i}{\tau^2} f(\nu_i,\mu_i)}}{w_1^{7/12} w_2^{7/12}}+\mathcal O(q^\frac{1}{2}), \qquad z_* \in S_\pm \setminus \{\pm1\}\ ,
\end{align}
Similarly, the leading term of $\mathcal H$ is also simplified at each BAE solution as follows:
\bea
&&  \mathcal H\biggr\vert_{z = z_*} \; =\;
q^\frac{7}{12}\times\frac{(3\mp1) e^{\frac{\pi i}{\tau} h(\nu_i,\mu_i)} (1-w_1^2) (1-w_2^2) (1-w_1 w_2)}{w_1^{3/2} w_2^{3/2}}+\mathcal O(q^\frac{13}{12})\ ,
\eea
where $ z_* \in S_\pm \setminus \{\pm1\}$ and we have used \eqref{eq:S-} for $z_* \in S_-$.

As a result, the generic leading term at $u_H/\tau = 1/2$ is given by
\bea
 \sum_{z_* \in S_\pm \setminus \{\pm1\}}  \mathcal F^p \, \mathcal H^{g-1}\biggr\vert_{z = z_*} &\sim&
q^{\frac{7}{12} (g-1)}\times  \left[2^g+(-1)^p 4^g\right] e^{\frac{\pi i}{\tau^2} p f(\nu_i,\mu_i)+\frac{\pi i}{\tau} (g-1) h(\nu_i,\mu_i)}\cr\cr
&&\times \frac{ \left[(1-w_1^2) (1-w_2^2) (1-w_1 w_2)\right]^{g-1}}{w_1^{\frac{7}{12} p+\frac{3}{2} (g-1)} w_2^{\frac{7}{12} p+\frac{3}{2} (g-1)}}\ .
\eea
Unless $g = 0$ and $p$ is odd, this term does not vanish, and the leading $q$-exponent is given by
\begin{align}
 \mathcal F^p \, \mathcal H^{g-1}\biggr\vert_{u_H/\tau = 1/2} \sim q^{\frac{7}{12} (g-1)}\ .
\end{align}
On the other hand, if $g = 0$ and $p$ is odd, this naive leading term cancels out. In such cases, we numerically find the true leading term, which turns out to be of order $q^\frac{5}{12}$.

At $u_H/\tau = 0$, on the other hand, the reduced BAE is given by
\begin{align}
\frac{\prod_{i = 1}^2 (z-y_i) \prod_{j = 1}^{2} (z^2-w_j)^2}{\prod_{i = 1}^2 (1-z y_i) \prod_{j = 1}^{2} (1-z^2 w_j)^2} = 1\ ,
\end{align}
with the anomaly-free condition $y_1 y_2 w_1^4 w_2^4 = 1$. Since it is difficult to solve this equation analytically, instead, we tried numerical analysis for given random phase values of $y_i, \, w_j$ and found
\begin{align}
 \mathcal F^p \, \mathcal H^{g-1}\biggr\vert_{u_H/\tau = 0} \sim q^{-\frac{5}{12} (g-1)}\ ,
\end{align}
which shows the exact agreement with the leading exponent at $u_H/\tau = 0$ predicted by \eqref{aH}. Thus, there is no cancelation of the leading terms at $u_H/\tau = 0$.

Combining these results at $u_H/\tau=0$ and at $u_H/\tau=1/2$, the leading term of
the total partition function is given by
\begin{align}
\Omega_{g,p} \sim \left\{\begin{array}{ll}
q^{-\frac{7}{12}}, \qquad & g = 0, \quad p \text{ even}, \\
q^{-\frac{5}{12} (g-1)}, \qquad & \text{otherwise.}
\end{array}\right.
\end{align}
for $SU(2)$ theory with two fundamental chirals and two
adjoint chirals.

Next, we move on to the physical gauge case. As advocated by Closset et.al. \cite{Closset:2017zgf},
the integer quantization condition for $R$-charges can now be relaxed,
and we can consider theories that flow to nontrivial superconformal points.
The superconformal $R$-charge is then determined by the anomaly-free
condition and the $a$-maximization. For physical gauge, a canonical
example with potential $H$-saddles is the ISS model,
which is the $SU(2)$ model with a single isospin-3/2 matter. From
the condition in section \ref{sec:locating H},
one can determine the $H$-saddles in the large radius limit as
\begin{align}
\hat u_H/\tau \quad = \quad \frac{6}{35}\,, \quad \frac{3}{10}\,, \quad \frac{1}{2}\,, \quad \frac{7}{10}\,, \quad \frac{29}{35} \ .
\end{align}
Note that for physical gauge in the large radius limit, BAE does depend on the manifold because it contains $l_R = \frac{1-g}{p}$. For simplicity we stick to $l_R = 1$ cases.

As we mentioned, for $p = 1, \, g = 0$, i.e., the superconformal index, the partition function can be written as the unit circle contour integral, which predicts the Casimir energy
\begin{align}
E_0 &=\sum_{\rho \in [3/2]}
\frac12 \left(\frac{(\epsilon'_\rho)^3}{3} +(\epsilon'_\rho)^2 m'_\rho-\epsilon'_\rho m'_\rho-\frac{\epsilon'_\rho}{6}+\frac{m'_\rho}{6}\right)\biggr\vert_{\epsilon'_\rho = \frac{3}{5}, m'_\rho = -1} \nonumber \\
&\quad +\sum_{\alpha \in [1]}
\frac12 \left(\frac{(\epsilon'_\alpha)^3}{3} +(\epsilon'_\alpha)^2 m'_\alpha-\epsilon'_\alpha m'_\alpha-\frac{\epsilon'_\alpha}{6}+\frac{m'_\alpha}{6}\right)\biggr\vert_{\epsilon'_\alpha = 0, m'_\alpha = 1} \nonumber \\
&= \frac{511}{1500}\ .
\end{align}
On the other hand, for each $H$-saddle, the reduced BAE and the leading terms of $(\mathcal F^\text{phys})^p H^{g-1}$ are given by
\begin{align}
\begin{array}{rl}
-z^7 = 1 \qquad & \text{at} \quad \hat u_H/\tau = \frac{6}{35} \ , \\
z^{-2} = 1 \qquad & \text{at} \quad \hat u_H/\tau = \frac{3}{10} \ , \\
z^8 = 1 \qquad & \text{at} \quad \hat u_H/\tau = \frac{1}{2} \ , \\
z^{-2} = 1 \qquad & \text{at} \quad \hat u_H/\tau = \frac{7}{10} \ , \\
-z^7 = 1 \qquad & \text{at} \quad \hat u_H/\tau = \frac{29}{35}\ ,
\end{array}
\end{align}
and
\begin{align}
\begin{array}{ll}
q^{-\frac{23}{10500}} \times \frac{1}{7 z^2} \qquad & \text{at} \quad \hat u_H/\tau = \frac{6}{35} \ , \\
q^{\frac{61}{1500}} \times \frac{z}{2} \qquad & \text{at} \quad \hat u_H/\tau = \frac{3}{10} \ , \\
q^{\frac{211}{1500}} \times \frac{(1-z^2)^{2}}{8 z^{6}} \qquad & \text{at} \quad \hat u_H/\tau = \frac{1}{2} \ , \\
q^{\frac{61}{1500}} \times \frac{z}{2} \qquad & \text{at} \quad \hat u_H/\tau = \frac{7}{10} \ , \\
q^{-\frac{23}{10500}} \times \left(-\frac{1}{7 z^5}\right) \qquad & \text{at} \quad \hat u_H/\tau = \frac{29}{35}\ ,
\end{array}
\end{align}
with $z = e^{2 \pi i \sigma}$. Note that $z = \pm 1$ at $\hat u_H/\tau = 1/2$ are Weyl invariant and again excluded from the solution set. One can see that those leading contributions all vanish for $p = 1$ as expected. We also confirmed numerically that the sub-leading terms with $q$-exponents less than $\frac{511}{1500}$ are all canceled out such that the true leading term of the total partition function is of order $q^\frac{511}{1500}$. Note that the value happens to coincide with the would-be exponent at $\hat u_H/\tau=0$, even though the tower sits at the $\hat u_H /\tau= 1/2$  saddle. This is one example of cancelations for $SCI$'s which was advertised previously.

On the other hand, such cancelations in favor of the would-be
exponent at $u_H=0$ does not necessarily happen for general values of $p$.
For $p = -2, \, g = 3$, as the second example, the leading $(\mathcal F^\text{phys})^p H^{g-1}$ is given by
\begin{align}
\begin{array}{ll}
q^{\frac{23}{5250}} \times 49 z^4 \qquad & \text{at} \quad \hat u_H/\tau = \frac{6}{35} \ , \\
q^{-\frac{61}{750}} \times \frac{4}{z^2} \qquad & \text{at} \quad \hat u_H/\tau = \frac{3}{10} \ , \\
q^{-\frac{211}{750}} \times \frac{64 z^{12}}{(1-z^2)^4} \qquad & \text{at} \quad \hat u_H/\tau = \frac{1}{2} \ , \\
q^{-\frac{61}{750}} \times \frac{4}{z^2} \qquad & \text{at} \quad \hat u_H/\tau = \frac{7}{10} \ , \\
q^{\frac{23}{5250}} \times 49 z^{10} \qquad & \text{at} \quad \hat u_H/\tau = \frac{29}{35}\ .
\end{array}
\end{align}
The locations of $H$-saddles are the same as those of $p = 1, \, g = 0$ because the two geometries share the same $l_R = 1$. Substituting the BAE solutions at each $H$-saddle, the leading terms are evaluated as
\begin{align}
\begin{array}{ll}
q^{\frac{23}{5250}} \times 0 \qquad & \text{at} \quad \hat u_H/\tau = \frac{6}{35} \ , \\
q^{-\frac{61}{750}} \times 8 \qquad & \text{at} \quad \hat u_H/\tau = \frac{3}{10} \ , \\
q^{-\frac{211}{750}} \times 72 \qquad & \text{at} \quad \hat u_H/\tau = \frac{1}{2} \ , \\
q^{-\frac{61}{750}} \times 8 \qquad & \text{at} \quad \hat u_H/\tau = \frac{7}{10} \ , \\
q^{\frac{23}{5250}} \times 0 \qquad & \text{at} \quad \hat u_H/\tau = \frac{29}{35}\ .
\end{array}
\end{align}
Thus, the leading $q$-exponent $-\frac{211}{750}$ persists
in this case, meaning that the nontrivial $\hat u_H/\tau=1/2$
saddle is dominant and the leading exponent there suffers
no cancelations, in contrast to the $p=1$ case.

\subsubsection*{Anomaly or Not}

Several observations relating these asymptotic coefficients
to the axial and to the conformal anomalies were made recently
\cite{DiPietro:2014bca,Bobev:2015kza,Martelli:2015kuk,Assel:2015nca,Hosseini:2016cyf}.
One well-known example is a relation between the Cardy exponent and
the sum of $U(1)_R$ charges of fermions, on par with (\ref{naiveCardy}),
which, for superconformal cases, translates to ``$a-c$" where
$a$ and $c$ are the usual conformal anomaly coefficients.

One main consequence of our investigation is that such
a connection cannot be trusted in general. Whenever the matter
content involves gauge representation beyond the simplest
ones, $H$-saddles will tend to appear at $u_H\neq 0$,
some of which could dominate the naive one at $u_H=0$
easily. The canonical example of SQCD escapes this,
since the chiral multiplets are all in the fundamental representation.
It probably explains why this rather generic phenomenon
has so far failed to be noticed. For SCI's, such
a deviation from $\frac{(1-g)}{12}\cdot {\rm tr}_f R$
has been observed first by Ardehali \cite{Ardehali:2015bla}
and subsequently by Di Pietro and Honda \cite{DiPietro:2016ond} for a handful
of examples, but, as we saw, this deviation is more of a
rule than an exception.

We also saw that something similar happens with the
Casimir limit as well. We again find that the notion
of $H$-saddles would be valid even in the large radius limit
provided that there are two circles in the spacetime,
at least for computation of the partition functions.
This will generally complicate the asymptotics of
typical partition functions, just as in the Cardy limit.
For this Casimir energy side, however, the connection to
the global anomaly \cite{Bobev:2015kza,Assel:2015nca} is
a little more robust than the Cardy side, although somewhat
dependent on the background geometry;  SCI's, in particular,
turned out to enjoy a rather special structure such that
this naive Casimir energy, apparently from $u_H=0$,
stands uncorrected even though nontrivial $H$-saddles exist,
and, more surprisingly, even when the naive $u_H=0$ saddle is absent,
due to magical cancelations between BAE vacua or even between
$H$-saddles. For general partition functions, for example with $p>1$,
such cancelations are more scarce.

Much of this  section explored such diverse forms
of the Casimir energies and the Cardy exponents, and
gave precise methods for isolating these, albeit with
no obvious universal formula.

\section{Summary}

We have introduced the notion of the holonomy saddle,
or $H$-saddle, and explored how the phenomenon
manifests in $d=4$ $\cN=1$ massive gauge theories.

Certain discrete values of the gauge holonomy are
found to support $d=3$ $\cN=2$ supersymmetric gauge
theories. When the space
is taken to be noncompact, the existence of multiple
$H$-saddles
means that a theory $G$ compactified on a small circle
admits multiple superselection
sectors at discrete holonomy values $u_H$'s,
where supersymmetric vacua are clustered which are
in turn attributable to an effective 3d theory $H$. Such an
$H$-theory tends to have
generally smaller light field content than the naive
dimensional reduction, due to the symmetry breaking by
the Wilson line, although one also typically finds
the naive saddle at $u_H=0$ as well. This observation
dovetails nicely against some of the existing studies of 4d-to-3d
and 3d-to-2d reduction of dualities \cite{Aharony:2013dha,Aharony:2017adm},
where one finds that a single dual pair typically
generates multiple dual pairs in the lower dimensions.

Its manifestation in the compact spacetime equipped
with a circle, on the other hand, implies precise
relations between the twisted partition function
of the $G$ theory and those of the subsequent $H$ theories,
to which we have devoted the bulk of the computations.
As such, the Witten index of the $G$ theory would
be generally a sum of Witten indices of the $H$ theories,
which explains, in part, how the number of the supersymmetric
vacua differs between two theories in the adjacent dimensions
even with the same supermultiplet content. This also offers
a definite method for reconstructing one from the others.

We also investigated the consequences for
supersymmetry-preserving torus-fibred compact spacetimes
by observing how the
twisted partition functions behave in small radius
limit of one of the two circle fibers. The  4d twisted
partition function reduces to a sum of 3d partition
functions in those limits, modulo exponential prefactors
which are interpreted either as the Cardy
or as the Casimir behavior, depending on which direction
is taken to be the Euclidean time. The results on such
exponents are generally different from the existing
claims, as the latters tend to focus, effectively, on the
naive saddle at $u_H=0$.

In the current examples of partition functions and
theories, which admit the BAE
description, $H$-saddles are located by asking which
subset of chiral matter fields become light at which
discrete values of the holonomy. FI constants and
Chern-Simons levels, generated by KK modes, can
further complicate the pattern, which we also
delineated in much detail.
The characterisation of $H$-saddles should be a
bit more general, however: an $H$-saddle would appear in the
holonomy space wherever the dimensionally reduced
theory admits supersymmetric vacua, normalizable or
non-normalizable \cite{Hwang:2017nop}. This general
criterion for $H$-saddles should be valid for any
superysmmetric gauge theories, as long as the gauge
holonomy is not exactly flat at the quantum level.

What we have not explored here is how this phenomenon
relates to and interacts with the matter of disconnected
holonomy sectors, well-known in the context of 4d
Witten index computations of pure Yang-Mills theories.
Since our $H$-saddles would occur already for the holonomy
on $S^1$ and since the corresponding discrete choices
$u_H$ arise from the dynamics rather than from the topology,
it is clear that the topological consideration must be separately
considered as well for more general gauge group $\cG$. An
immediate question is how the discussion here should
be generalized when $\cG$ is not simply connected \cite{Witten:2000nv}
or when the so-called ``triple" is relevant
\cite{Witten:1997bs,Keurentjes:1999mv,Keurentjes:1999qf,Kac:1999gw}.
We suspect we will encounter more issues related to such
holonomy saddles and holonomy islands in near future.

\vskip 0.5cm

\section*{Acknowledgement}
We would like to thank Cyril Closset, Richard Eager,
Heeyeon Kim, Nati Seiberg, and Edward Witten for
useful conversations. The research of S.L. is supported in part by the National
Research Foundation of Korea (NRF) Grant NRF-2017R1C1B1011440.

\end{document}